\documentclass[aip,jcp,longbibliography
]{revtex4-2}
\usepackage{bm}
\usepackage{physics}
\usepackage{qcircuit}
\usepackage{graphicx}
\usepackage{color}
\usepackage{float}
\usepackage{array, multirow}
\usepackage{amsmath}
\usepackage{xcolor}
\usepackage{booktabs}
\usepackage{soul}
\usepackage{url}
\usepackage{hyperref}
\usepackage{newtxtext}
\usepackage{amssymb}

\draft 

\begin{document}


\title{
Unitary Dynamics for Open Quantum Systems with Density-Matrix Purification
} 



\author{Luis H. Delgado-Granados}
\affiliation{Department of Chemistry and The James Franck Institute, The University of Chicago, Chicago, IL 60637 USA}
\author{Samuel Warren}
\affiliation{Department of Chemistry and The James Franck Institute, The University of Chicago, Chicago, IL 60637 USA}
\author{David A. Mazziotti}
\email[]{damazz@uchicago.edu}
\affiliation{Department of Chemistry and The James Franck Institute, The University of Chicago, Chicago, IL 60637 USA}


\date{Submitted May 29, 2024}

\begin{abstract}
Accurate modeling of quantum systems interacting with environments requires addressing non-unitary dynamics, which significantly complicates computational approaches. In this work, we enhance an open quantum system (OQS) theory using density-matrix purification, enabling a unitary description of dynamics by entangling the system with an environment of equal dimension. We first establish the connection between density-matrix purification and conventional OQS methods. We then demonstrate the standalone applicability of purification theory by deriving system-environment interactions from fundamental design principles. Using model systems, we show that the purification approach extends beyond the complete positivity condition and effectively models both Markovian and non-Markovian dynamics. Finally, we implement density-matrix purification on a quantum simulator, illustrating its capability to map non-unitary OQS dynamics onto a unitary framework suitable for quantum computers.
\end{abstract}


\maketitle 


\section{Introduction}

The prediction of how a quantum system changes upon its interaction with an environment is fundamental to a wide range of fields, arising in biological systems, such as in the study of light-harvesting systems~\cite{mohseni2014quantum},
or in the development of novel quantum technologies, such as in the search for new qubit candidates~\cite{doi:10.1021/acs.chemrev.0c00620}. To facilitate such studies, the theory of open quantum systems (OQS) was developed to model the dynamics of an initially uncorrelated system in the context of its interactions with the environment~\cite{Breuer2002, jagadish2019, manzano2020,breuer2016}.

Many techniques have been developed to model OQS, such as hierarchical equations of motion (HEOM)~\cite{10.1063/5.0011599, 10.1063/5.0082822,Chan_2018}, tensor-based methods~\cite{strathearn2018efficient,e22090984,Lyu.2023,Clark_2010,JASCHKE201859}, and even machine learning approaches using restricted Boltzmann machines (RBM)~\cite{vicentini2019variational,PhysRevB.99.214306,PhysRevLett.122.250502,doi:10.1126/science.aag2302}. Although these approaches move towards a more general study of OQS by including non-Markovian behavior (e.g., HEOM)~\cite{10.1063/5.0011599, 10.1063/5.0082822,Chan_2018}, the question of how to model a system that is initially correlated with its environment remains open~\cite{breuer2016}. In addition, the emergence of quantum computing opens new opportunities and challenges for the simulation of OQS~\cite{Bacon.2001,Müller.2011,Wang.2011,Mostame.2012,Sweke.2015,Mostame.2016,Sweke.2016,Suri.2018,Inoue.2018,Gupta.2020,García-Pérez.2020,Endo.2020,Patsch.2020,Gupta.2020xew,Hu.2020,Lin.2021,Lee.2021,Schlimgen.2021,Head-Marsden.202180m,Gaikwad.2022,Jong.2022,Kamakari.2022,Tornow.2022,Schlimgen.2022,Schlimgen.2022lz7,Liang.2023,Lau.2023,Wang.20239eb,Li.2023,Miessen.2023,Peetz.2023,Régent.2023,Schlegel.2023,Zhou.2023nw3,Rossini.2023,David.2023,Suri.2023,Zhang.2023z47,Avdic.2023,Guimarães.2023,Santos.2024,Guimarães.2024,Luo.2024,Re.2024,Kunold.2024,Chen.2024,Shivpuje.2024,Basile.2024,Seneviratne.2024,Ding.2024,Watad.2024}.  Quantum computing has the potential to significantly reduce the computational cost of simulating OQS; at the same time, it raises the challenge of treating the non-unitary dynamics of OQS within the unitary framework of universal quantum computers.


Here, we further pursue an alternative approach to OQS based on density-matrix purification~\cite{schlimgen2022, 10.1063/1.5121749,PhysRevLett.129.066401,doi:10.1142/S0217979296000817,wilde2013,nielsen2010,Inoue.2018,viennot2018, PhysRevA.73.062309}, which recovers a unitary description of dynamics by entangling a maximally mixed system with an environment of equal dimension as the system. First, we connect OQS to its conventional theoretical framework based on quantum channels.  The density-matrix purification, we show, provides a significant generalization of quantum channels.  Second, we present a set of principles for designing the system-bath interaction to treat Markovian and non-Markovian dynamics.  Importantly, the density-matrix purification can treat cases beyond the restrictions of complete positivity. Finally, we demonstrate the theory through a quantum simulation, highlighting that the purification approach provides a unitary framework that can facilitate the simulation of OQS on quantum devices.


\section{Theory}
\label{Theory}

In section~\ref{sec:dmp} we discuss density-matrix purification with connections to the conventional approach of quantum channels, and in section~\ref{sec:eom} we derive an explicit relation between the treatment of OQS by the Lindblad equation and its treatment by density-matrix purification.

\subsection{Density-matrix purification}

\label{sec:dmp}

In an open quantum system (OQS), the state of the quantum system $S$ and its environment $E$, to which $S$ is coupled, at a time $t$ can be expressed by a density matrix $\rho_{SE}(t)$ which belongs to the Hilbert space of the composite system $\mathcal{H}_{SE}=\mathcal{H}_S \otimes \mathcal{H}_E$ ~\cite{Breuer2002,rivas2012open}. By tracing out the environment from $\rho_{SE}(t)$ and using the spectral expansion, we obtain the reduced density matrix of the OQS
\begin{align}
\label{eq:rdm}
\rho_{S}(t)&=\text{Tr}_E\left(\rho_{SE}(t)\right)\\
&=\sum_i \omega_i(t)\ket{\Psi^S_i(t)}\bra{\Psi^S_i(t)},
\end{align}
\noindent where $\omega_i(t)$ and $\ket{\Psi^S_i(t)}$ are the eigenvalues and eigenstates of the system $S$ at time $t$, respectively~\cite{Breuer2002,rivas2012open}.

Two common methods for describing the dynamics of $\rho_S(t)$ are: Quantum maps $\mathcal{E}$ and quantum master equations (QME)~\cite{Breuer2002, mohseni2014quantum, jagadish2019}. Both rely on the assumption that the system and environment are initially uncorrelated. Although the assumption of an initially uncorrelated system and environment is widely used, it limits the dynamics that can be studied because some physical processes do not possess this property~\cite{jordan2004, Rivas2014}. Mathematically, the system's embedding in the environment can be represented by an extension or assignment map $E_v$ ~\cite{pechukas1994, carteret2008}. For an initially uncorrelated system and environment, $E_v$ is a direct tensor product between the initial states of the system and the environment~\cite{pechukas1994, Breuer2002}
\begin{equation}
\label{eq:em}
E_v: \rho_S(t)\rightarrow\rho_S(t)\otimes\rho_E(t).
\end{equation}
Neglecting the initial interactions between the system and environment causes the dynamics to be linear, Hermitian, trace-preserving, and completely positive (CP). CP, often taken as a necessary condition to model physical dynamics of OQS~\cite{jagadish2019, manzano2020}, has been heavily debated due to its reliance on an initially uncorrelated system and environment, which leads to $E_v$ defined in Eq.~(\ref{eq:em})~\cite{pechukas1994, alicki1995, pechukas1995, carteret2008, sargolzahi2020, breuer2016}.

To apply the purification theorem, one introduces an external system $B$, which can be interpreted as an effective environment for $S$. The system $S$ is then entangled to the effective bath $B$ ~\cite{schlimgen2022, wilde2013, nielsen2010}. Just as with quantum maps and QMEs, the purification approach can be expressed as an extension map $E_v$
\begin{equation}
\label{eq:emp}
E_v: \rho_S(t)\rightarrow\ket{\Psi^{SB}(t)}\bra{\Psi^{SB}(t)},
\end{equation}
where
\begin{equation}
\label{eq:pw}
\ket{\Psi^{SB}(t)}=\sum_i \sqrt{\omega_i(t)}\ket{\Psi^S_i(t)}\ket{\Psi^B_i(t)},
\end{equation}
in which $\ket{\Psi^{SB}(t)}$ is a pure state and $\{\ket{\Psi^B_i(t)}\}$ is a complete orthonormal basis set for $B$ $\left(\ket{\Psi^B_i(t)}\in \mathcal{H}_{B}\right)$~\cite{schlimgen2022, wilde2013, nielsen2010}. By defining the new composite system as $SB$, the density-matrix purification allows for a unitary description of the dynamics~\cite{Breuer2002,schlimgen2022}:
\begin{align}
\label{eq:upw}
\ket{\Psi^{SB}(t')} &= \hat{U}_{SB}(t',t)\ket{\Psi^{SB}(t)}\\
&={\hat T} \exp{-i\int_{t}^{t'} \hat{H}_T(\tau) d\tau}\ket{\Psi^{SB}(t)},
\end{align}
where ${\hat T}$ is the time-ordering operator and $\hat{H}_T$ is total Hamiltonian of $SB$
\begin{equation}
\label{eq:ht}
\hat{H}_T(t)=\hat{H}_{S}(t)\otimes\hat{I}_B+\hat{I}_{S}\otimes\hat{H}_B(t) + \hat{H}_{SB}(t)
\end{equation}
where $\hat{H}_{S}(t)$, $\hat{H}_{B}(t)$, and $\hat{H}_{SB}(t)$ are the Hamiltonians of the system,  the effective environment, and the interaction between the two, respectively~\cite{schlimgen2022}.

The purification approach does not specify how one defines $B$, the selection of $\{\ket{\Psi^B_i(t)}\}$ or the dimension of $\mathcal{H}_{B}$, dim$(\mathcal{H}_{B})$. The only condition imposed is that dim$(\mathcal{H}_{B})$ is bounded from below by the rank of $\rho_S(t)$, which in the case of a maximally mixed $\rho_S(t)$ equals the system's dimension $d$~\cite{wilde2013, nielsen2010}. This is in direct contrast with the conventional approaches to OQS, in which, as described by the \textit{Stinespring's Theorem}~\cite{Suri2023,ticozzi2017quantum}, the environmental dimensions are bounded from above by $d^2$~\cite{Suri2023,ticozzi2017quantum}. The decreased size of the effective bath in the purification method may give an advantage over conventional approaches to OQS by decreasing the simulation's computational cost on both classical and quantum devices~\cite{schlimgen2022}.

As can be seen by contrasting Eq.~(\ref{eq:em}) and Eq.~(\ref{eq:emp}), the embedding of the purification method differs from the conventional approach, not only by allowing initial correlations between system systems and environment but also behaving non-linearly. The flexibility of the purification approach to capture a broad range of dynamics of OQS (eg., non-Markovian and non-CP dynamics), as we show in this work, can be related to the non-linearity in the preparation of $\ket{\Psi^{SB}(t)}$. $E_v$ allows us to connect different sub-spaces that otherwise would not be accessible to each other~\cite{ando1986}.


\subsection{Equation of motion}

\label{sec:eom}

In this work, we examine the construction of $\hat{H}_{SB}(t)$, which has not been attempted from the purification approach perspective.  We use the differential representation of the dynamics of $SB$ described by the density-matrix purification approach, which can be derived from Lioville's equation
\begin{equation}
\label{eq:lioeq}
\dot{\rho}_S(t)=-i\text{Tr}_B\left( \left[\hat{H}_T(t),E_v(\rho_S(t))\right] \right),
\end{equation}
where we have traced over the degrees of freedom of the effective bath. By replacing $\hat{H}_T$ with the Hamiltonian in Eq.~(\ref{eq:lioeq}), we obtain three terms
\begin{equation}
\begin{aligned}
\label{eq:lioeq-rhs}
\dot{\rho}_S(t)=&-i \text{Tr}_B\left(\left[\hat{H}_S\otimes \hat{I}_B,E_v(\rho_S(t))\right]\right)\\
&-i \text{Tr}_B\left(\left[\hat{I}_S \otimes\hat{H}_B,E_v(\rho_S(t))\right]\right)\\
&-i \text{Tr}_B\left(\left[\hat{H}_{SB},E_v(\rho_S(t))\right]\right).
\end{aligned}
\end{equation}
Tracing over the effective bath degrees of freedom using
\begin{equation}
\text{Tr}_B\left(\bullet\right)=\sum_k\bra{\Psi^B_k(t)}\bullet\ket{\Psi^B_k(t)},
\end{equation}
we obtain
\begin{equation}
\begin{aligned}
\label{eq:lioeq-rhs-s}
\dot{\rho}_S(t)=&-i \left[\hat{H}_S,\rho_S(t)\right]\\
&-i \text{Tr}_B\left(\left[\hat{H}_{SB},E_v(\rho_S(t))\right]\right),
\end{aligned}
\end{equation}
in which the first term describes the dynamics of the system generated by $\hat{H}_S$, and the second term represents the dissipation due to the interaction between the system and the effective environment, $\hat{H}_{SB}$. The reduction in the number of terms on the right-hand side (RHS) of Eq.~(\ref{eq:lioeq-rhs}) occurs because $\hat{H}_B$ act only on the bath and therefore, does not contribute to the system's dynamics.

Eq.~(\ref{eq:lioeq-rhs-s}) resembles the structure of the the Gorini-Kossakowski-Sudarshan-Lindblad (GKSL) QME~\cite{Breuer2002, mohseni2014quantum}:
\begin{align}
\label{eq:le}
\dot{\rho}_S(t) &= -i\left[\hat{H},\rho_S\right] +\sum_{k=1}\gamma_k\left(\hat{L}_k\rho_S\hat{L}^\dagger_k+\frac{1}{2}\{\hat{L}^\dagger_k\hat{L}_k,\rho_S\}\right)\\
& = -i\left[\hat{H},\rho_S\right] + \hat{\mathcal{D}}({\rho}_S(t))
\end{align}
where the first term of Eq.~(\ref{eq:le}) represents the unitary evolution of the dynamics of the system generated by the Hamiltonian $\hat{H}$~\cite{Breuer2002}. The second term, which makes the dynamics deviate from unitary, is called the \textit{dissipator} $\hat{\mathcal{D}}({\rho}_S(t))$~\cite{Breuer2002}. This term contains what are known as the Lindblad operators $\hat{L}_k$, and the relaxation rates for each mode of dissipation $\gamma_k$. In this equation, $\left[\bullet\right]$ $(\{\bullet\})$ is the (anti-)commutator. Even with these similarities, Eq.~(\ref{eq:lioeq-rhs}) and Eq.~(\ref{eq:le}) are fundamentally different since with the purification approach: (a) the dynamics are not restricted to initial states without system-environment interactions ($E_v$ is given by purification technique), (b) there is no weak-interaction approximation (Markovian approximation), and (c) Eq.~(\ref{eq:lioeq-rhs-s}) is non-linear.

As implied by Eq.~(\ref{eq:lioeq-rhs-s}), an important aspect of the density-matrix purification technique is building an interaction operator between the system and the effective environment, $\hat{H}_{SB}$. As shown in the next section, we can obtain an $\hat{H}_{SB}$ that successfully replicates Markovian dynamics, which are described by a linear CP quantum map. To achieve this, we equate the GKSL equation Eq.~(\ref{eq:le}) to Eq.~(\ref{eq:lioeq-rhs-s}), and set $\hat{H}$ to $\hat{H}_S$ in Eq.~(\ref{eq:le}), yielding
\begin{equation}
\label{eq:leq=lioeq}
\hat{\mathcal{D}}({\rho}_S(t))=
-i\text{Tr}_B\left( \left[\hat{H}_{SB}(t),E_v(\rho_S(t))\right]\right),
\end{equation}
which defines the behavior of the interaction term, $\hat{H}_{SB}(t)$, in terms of $\hat{L}_k$. In addition to replicating Markovian dynamics, as we mentioned before, the purification technique can describe non-Markovian dynamics and non-CP dynamics. This is done using a systematic approach similar to the one used in the field of coherent control of chemical reactions. The systematic approach consists of the heuristic search of different \textit{shapes} of interactions (e.g., Gaussian, periodic, etc.). The examples shown in the next section highlight the flexibility of the density-matrix purification technique and reinforce the advantages of having a non-linear extension map $E_v$ (see Eq.~(\ref{eq:lioeq-rhs-s})), as well as the reduced dimensionality of the effective bath compared to the conventional approaches.


\section{Results \& Discussion}

This section has three parts: \textit{Conventional approach} in section~\ref{sec:ca}, where we present how to replicate Markovian dynamics with the purification technique effectively, \textit{Beyond the traditional approach} in section~\ref{sec:bta}, in which we show how the purification technique leads to a natural way of applying the purification to model linear non-CP, Markovian and non-Markovian dynamics, and \textit{Simulating OQS in quantum computing} in section~\ref{sec:sqc}, where we demonstrate the use the purification technique to simulate non-Markovian dynamics on a quantum computing framework.

\subsection{Conventional approach}
\label{sec:ca}
To validate the connection between the Lindblad master equation and the purification approach shown in Eq.~(\ref{eq:leq=lioeq}), we study the decay channel for a two-level system. For this system, the Lindblad master equation is
\begin{equation}
\label{eq:le-decay}
\dot{\rho}_S(t) = -i\left[\hat{H},\rho_S\right] +\gamma_D\left(\hat{L}_D\rho_S\hat{L}^\dagger_D+\frac{1}{2}\{\hat{L}^\dagger_D\hat{L}_D,\rho_S\}\right),
\end{equation}
where
\begin{equation}
\label{eq:ham_2level}
\hat{H}=
\begin{bmatrix}
0 & 0 \\
0 & 1 \\
\end{bmatrix},
\end{equation}
and
\begin{equation}
\label{eq:L_op}
\hat{L}_D=
\begin{bmatrix}
0 & 1 \\
1 & 0 \\
\end{bmatrix}.
\end{equation}
and for this work $\gamma_D$ is $0.1$.

\begin{figure}[!ht]
    \centering
    \includegraphics[width=1\textwidth]{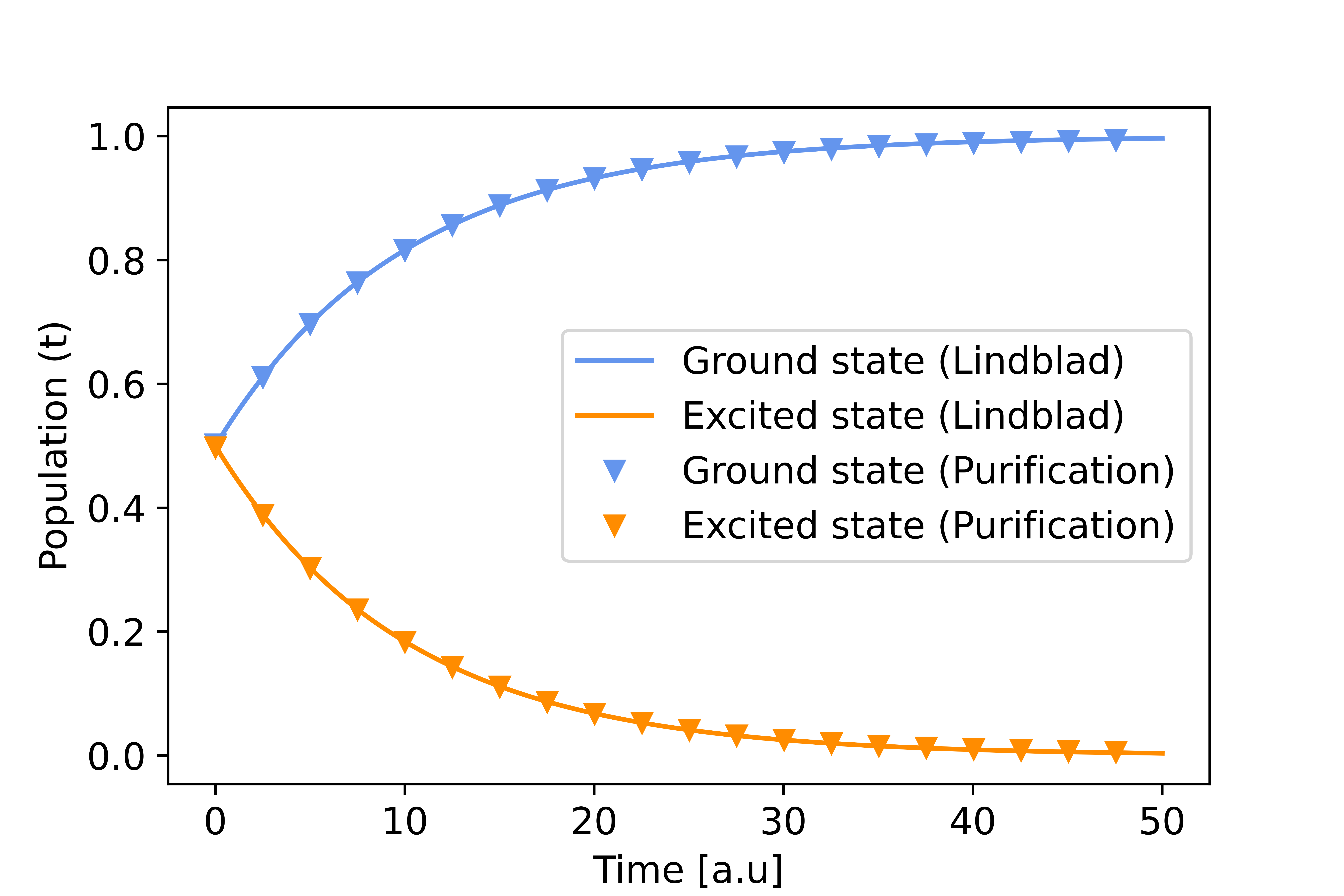}
     \caption{Dynamics of the two-level system following the Lindblad master equation (see Eq.~(\ref{eq:le-decay})) and the Purification approach (see Eq.~(\ref{eq:lioeq})).\label{fig:dynamics}}
\end{figure}

By using Eq.~(\ref{eq:leq=lioeq}), we are able to find an interaction Hamiltonian between the system and bath $\hat{H}_{SB}(t)$ that reproduces the Lindbladian dynamics when $\rho_S(0)=\frac{1}{2}(\ket{0}\bra{0}+\ket{1}\bra{1})$ (see Eq.~(\ref{eq:le-decay}) and Fig.~\ref{fig:dynamics}). The search for $\hat{H}_{SB}(t)$ is accomplished using the Scipy L-BFGS-B optimization method and, for convenience, by \textit{purifying} the state of the system with a copy of itself and creating an entangled state (see Eq.~\ref{eq:pw}). This is performed for different mixed states, showing the capabilities of the purification method. Being able to obtain the interaction term $\hat{H}_{SB}$ leads directly to a unitary description of the dynamics (see Eq.~(\ref{eq:upw})) that describes the system's dynamics at any point in time without having to perform additional dilations of the system.

Although this shows the potential of the purification theorem to replicate Markovian and CP dynamics with a unitary description, the interaction Hamiltonian's $\hat{H}_{SB}(t)$ behavior is nontrivial. This leads to $\hat{H}_{SB}(t)$ having a dependence on the initial state of the system and requiring one to work with mixed states.

\subsection{Beyond the traditional approach}
\label{sec:bta}
\subsubsection{Non-CP dynamics}

To show the flexibility of the purification technique in the conventional framework of OQS, we replicate the dynamics of the non-CP quantum map, $\mathcal{E}$, that maps the Bloch ball onto a unit disc~\cite{jagadish2019}
\begin{align}
\label{eq:ncp_map_0}
\mathcal{E}(\hat{I})&=\hat{I}\\
\label{eq:ncp_map_1}
\mathcal{E}(\hat{\sigma}_1)&=\hat{\sigma}_1\\
\label{eq:ncp_map_2}
\mathcal{E}(\hat{\sigma}_2)&=\hat{\sigma}_2\\
\label{eq:ncp_map_3}
\mathcal{E}(\hat{\sigma}_3)&=0,
\end{align}
where $\{\hat{\sigma}_{i}\}_{i=1,2,3}$ are the Pauli matrices acting on $S$.

\begin{figure}[!ht]
    \centering
    \includegraphics[width=1\textwidth]{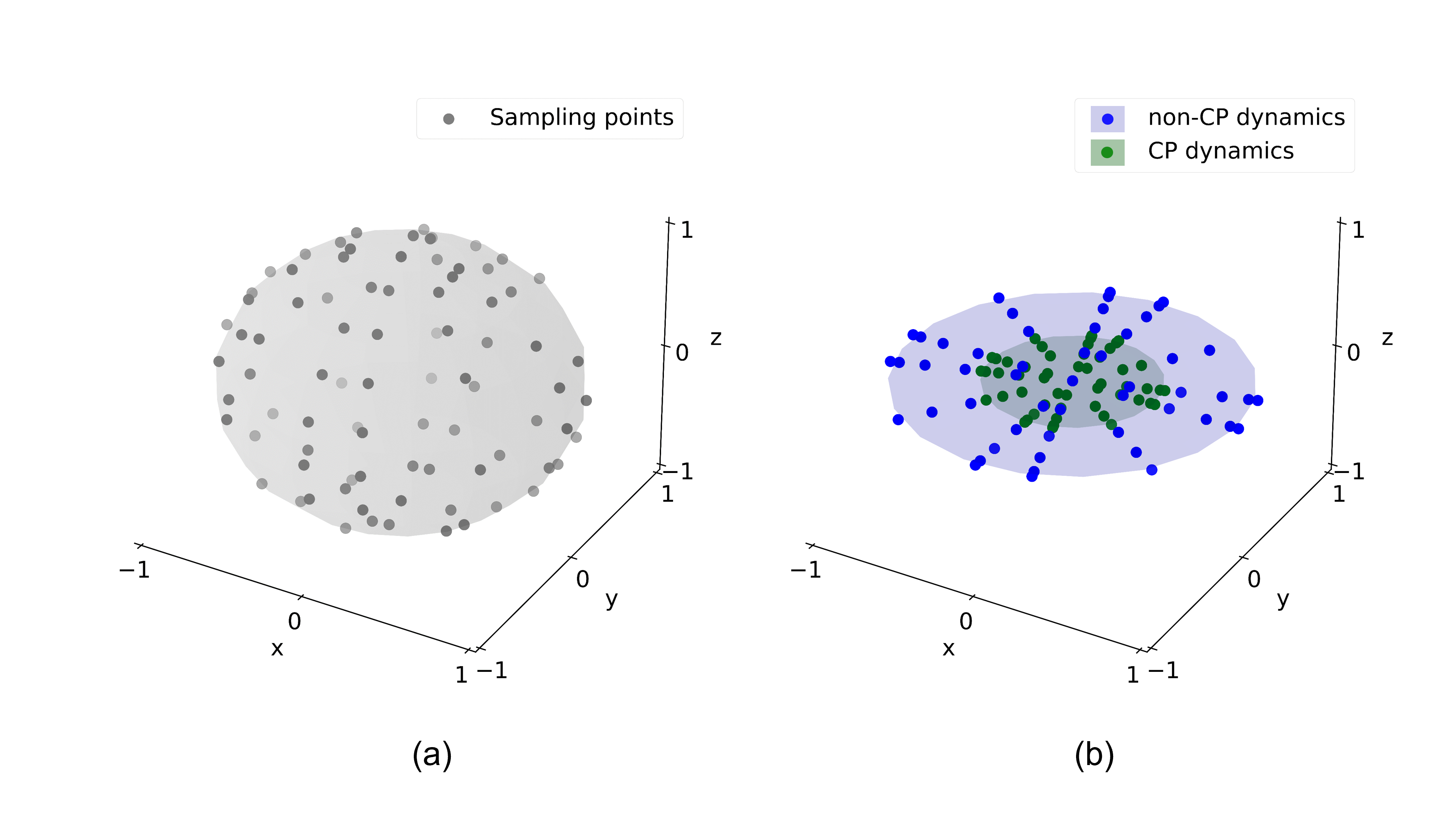}
     \caption{(a) Sampling done in the Bloch sphere. (b) Non-CP dynamics described by the unitaries obtained with Eq.~(\ref{eq:u_pur}) in contrast with the ones described by constraining $\mathcal{E}$ to be CP~\cite{jagadish2019}.\label{fig:dynamics_ncp}}

\end{figure}

For this case, we sample the Bloch sphere, as shown in Figure \ref{fig:dynamics_ncp}, and determine the action of the quantum map $\mathcal{E}$ (Eq.~(\ref{eq:ncp_map_0})-Eq.~(\ref{eq:ncp_map_3})) for each state.  To \textit{purify} each of the states, we consider the effective bath as a copy of itself; moreover, we define the unitary that connects them by
 \begin{equation}
\label{eq:u_pur}
\hat{U}_{NCP}=\mathbf{V}_T\mathbf{V}_S^\dagger,
\end{equation}
where $\mathbf{V}_S$ and $\mathbf{V}_T$ are matrices in which the eigenvectors of the \text{purified} initial and final state, respectively, are the columns. As shown in Fig.~\ref{fig:dynamics_ncp}, by using Eq.~(\ref{eq:u_pur}) for each sampled point, we are able to obtain a set of unitaries $\{U_{NCP}\}$ that map each sampled point into the unit plane represented by the blue disc. To show the contrast between non-CP and CP dynamics, we also plot the projection of the map that generates a disc but is restricted by CP, as shown by the green disc in Fig. \ref{fig:dynamics_ncp}~\cite{jagadish2019}. The projection, shown in Fig. \ref{fig:dynamics_ncp}, agrees with the data in Ref.~\citenum{jagadish2019}. These results emphasize the versatility of the density-matrix purification technique, taking us beyond the conventional approach, which is restricted by CP dynamics.


\subsubsection{Two-level system decay}

\begin{figure}[!ht]
    \centering    \includegraphics[width=1\textwidth]{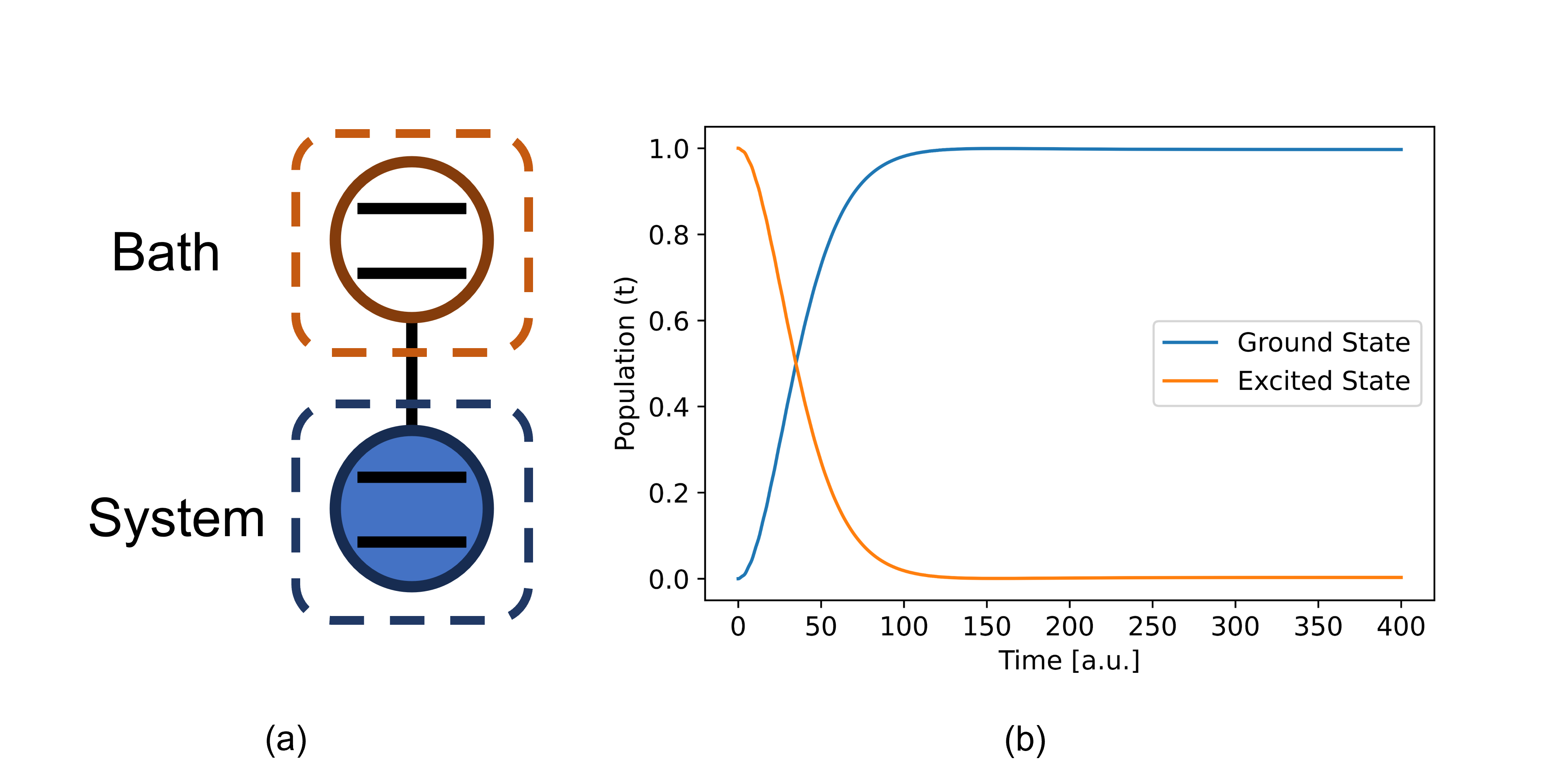}
     \caption{(a) Graph representation of $S+B$. (b) Population dynamics of $S$ showing the decay process.\label{fig:dynamics_2q}}
\end{figure}

To establish how one can use the density-matrix purification technique as a stand-alone method to model OQS, we start with the task of modeling a two-level system decay. We choose a two-level system for $S$ and $B$, both sharing the same Hamiltonian:
\begin{equation}
\label{eq:H_pur_2s}
\hat{H}_S=\hat{H}_B=
\begin{bmatrix}
0 & 0\\
0 & 1\\
\end{bmatrix}.
\end{equation}
To describe a decay process, we consider our system in its excited state with the effective bath in its ground state
\begin{equation}
\label{eq:rho_s_pur_2s}
\rho_S(t_0)=\ket{1}\bra{1}
\end{equation}
and
\begin{equation}
\label{eq:rho_b_pur_2s}
\rho_B(t_0)=\ket{0}\bra{0},
\end{equation}
where $\ket{0}=\begin{bmatrix}1&0\end{bmatrix}^T$ and $\ket{1}=\begin{bmatrix}0&1\end{bmatrix}^T$. We consider $B$ as an \textit{empty} bath that will be capable of receiving the full excitation. This description of $SB$ can be represented naturally by using a graph as shown in Fig.~\ref{fig:dynamics_2q}, in which each node in the graph represents a two-level system, the population in each system is given by its filling, and $\hat{H}_{SB}(t)$ is represented by the edge between the nodes $S$ and $B$.

Since $S$ and $B$ are degenerate, the interaction must be time-dependent to avoid an oscillatory behavior.  In our case, to define the interaction $\hat{H}_{SB}(t)$, we opt for a time-dependent function that physically represents decay, an exponential
\begin{equation}
\label{eq:H_sb_pur_2s}
\hat{H}_{SB}(t)=\exp{-\alpha t}\hat{H}'_{SB},
\end{equation}
where $\alpha\in\mathbb{R}$ and $\hat{H}'_{SB}$ is a Hermitian matrix acting on $\mathcal{H}_{SB}=\mathcal{H}_{S}\otimes\mathcal{H}_{B}$. To make the initial selection of $\alpha$ and the elements of $\hat{H}'_{SB}$, we perform an L-BFGS-B optimization with Scipy. The purpose of the optimization is to obtain parameters that lead to the system being in the ground state, $\rho_S(t_c)=\ket{0}\bra{0}$, at a specific time $t_c$. As shown in Fig.~(\ref{fig:dynamics_2q}), this yields an interaction $\hat{H}_{SB}(t)$ that correctly describes the decay process of $S$ assisted by $B$.

The behavior shown in Fig.~(\ref{fig:dynamics_2q}) reflects the decay of the interaction Hamiltonian due to the exponential term in Eq.~(\ref{eq:H_sb_pur_2s}), indicating that one can easily manipulate the rate of decay by modifying the parameter $\alpha$. Eq.~(\ref{eq:H_sb_pur_2s}) is not the only \textit{form} of the interaction that leads to a decay behaviour; in fact, a train of Gaussians
\begin{equation}
\label{eq:gauss_i}
\sum_i{\frac{a_i}{b_i\sqrt{2\pi}}\exp{-\frac{1}{2}\left(\frac{t-t_i}{b_i}\right)^2}},
\end{equation}
where $\{t_i,a_i,b_i\}\in\mathbb{R}$, or
\begin{equation}
\label{eq:sin_i}
\sin{(t^2+t)}^2
\end{equation}
can also describe a decay behavior in the system. This showcases how diverse the purification technique can be, not restricting itself to a specific type of interaction, but providing a pool of functions to replicate complex dynamics. In addition to this, the density-matrix purification technique does not put any restriction on the selection of the basis for $B$ or $\hat{H}_B$. This gives more flexibility and also control over how the transfer can be performed.

\subsubsection{Two-level system network}

We also study a network of two coupled two-level systems with the aim of generating a controlled decay of the excitation in the two-level system network. This behavior is relevant since it is the basis behind the energy transfer process of important physical systems such as the light-harvesting Fenna-Matthews-Olson (FMO) complex~\cite{cook2022quantum, mohseni2014quantum, engel2007evidence,Schouten.2023cha}.

\begin{figure}[!ht]
    \centering
    \includegraphics[width=1.1\textwidth]{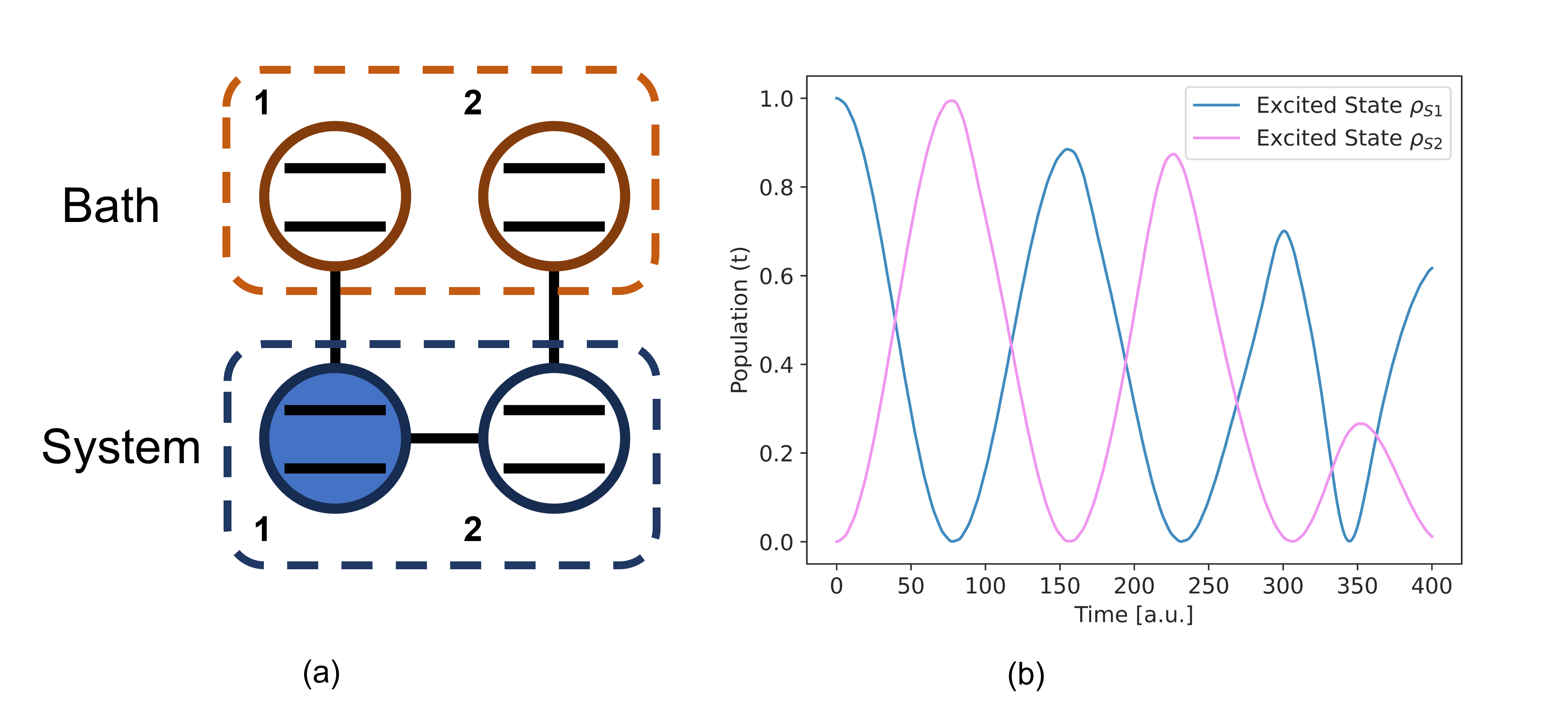}
     \caption{(a) Graph representation of $S+B$. (b) Population dynamics of the excited states of $S1$ and $S2$.\label{fig:dynamics_4q}}
\end{figure}

To \textit{purify} the system, we select a network of two-level systems for the bath, although the bath systems are not coupled. Just as in the previous case, we represent this system with a schematic graph as shown in Fig.~\ref{fig:dynamics_4q}, where each edge represents the interaction with the respective nodes and the filling of a node represents the population in the excited state of that sub-system. In this work, the Hamiltonian of $S$ is defined as
\begin{align}
\label{eq:h_s_4q}
\hat{H}_S&= \hat{H}_{S1}+\hat{H}_{S2}+\hat{C}_{S1,S2}\\
&=
\begin{bmatrix}
E_0 & 0\\
0 & E_1
\end{bmatrix}
+
\begin{bmatrix}
E_0 & 0\\
0 & E_1
\end{bmatrix}
+
\begin{bmatrix}
0 & C\\
C & 0
\end{bmatrix}\\
&=\begin{bmatrix}
2 E_0 & C\\
C & 2 E_1
\end{bmatrix}
\end{align}
and for $B$, we have that
\begin{align}
\label{eq:h_b_4q}
\hat{H}_B &= \hat{H}_{B1}+\hat{H}_{B2}\\
 &=
\begin{bmatrix}
E_0 & 0\\
0 & E_1
\end{bmatrix}
+
\begin{bmatrix}
E_0 & 0\\
0 & E_1
\end{bmatrix}\\
&=
\begin{bmatrix}
2 E_0 & 0\\
0 & 2 E_1
\end{bmatrix}
\end{align}
where $E_0=-0.5$, $E_1=0.5$, and $C=0.2$.

\begin{figure}[!ht]
    \centering
    \includegraphics[width=1\textwidth]{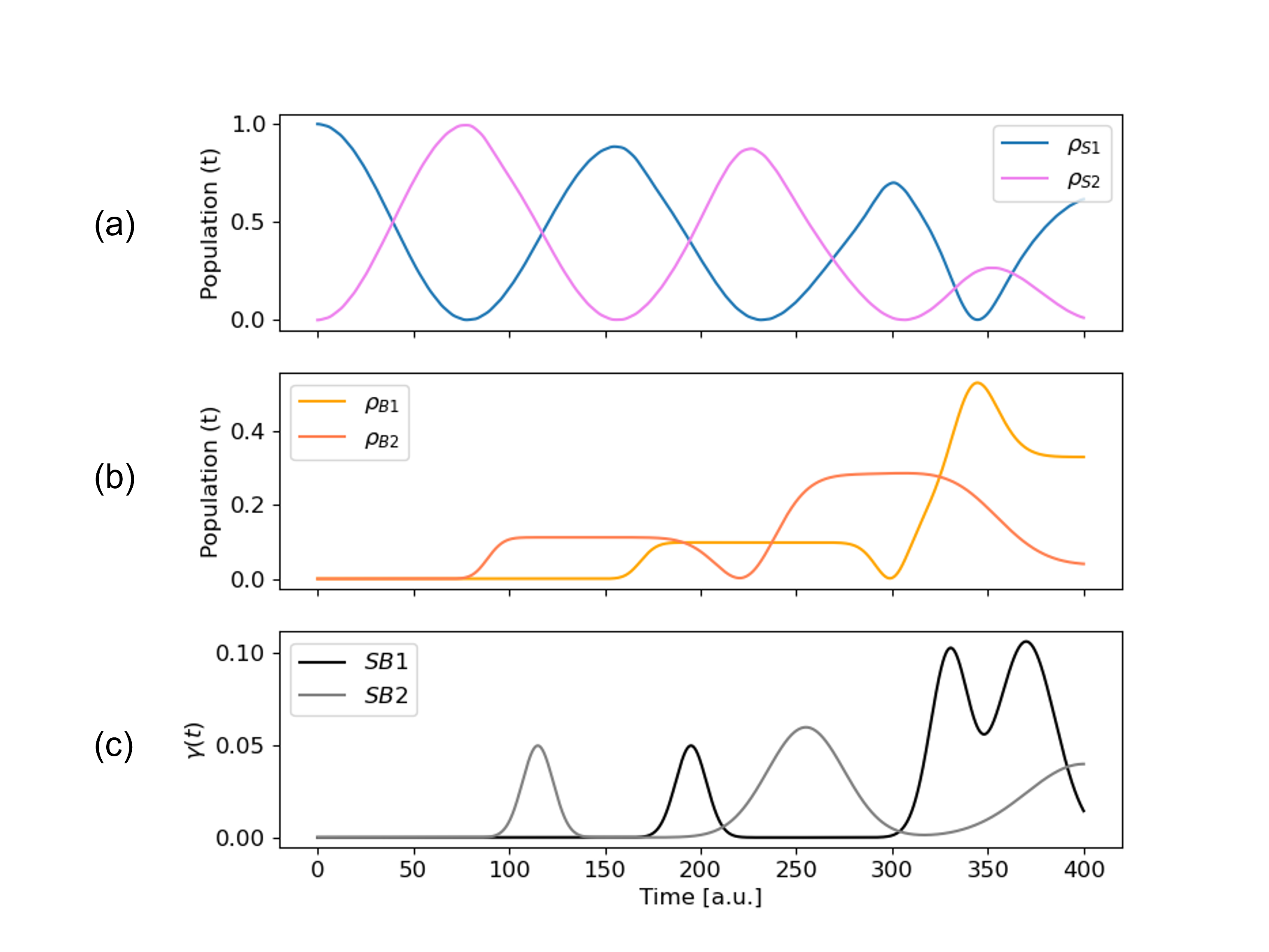}
     \caption{Population dynamics of the excited states for: (a) $S1$ and $S2$, (b) $B1$ and $B2$; as well as $\gamma(t)$ for both interactions, $\hat{H}_{SB1}(t)$ and $\hat{H}_{SB2}(t)$ in (c).
    \label{fig:dynamics_4q_sbi}}
\end{figure}

To generate the decay behavior, we heuristically search for the correct \textit{form} of $\hat{H}_{SB1}(t)$ and $\hat{H}_{SB2}(t)$,
\begin{equation}
\hat{H}_{SBi}(t)=\gamma(t)\hat{H}'_{SB}\quad i=\{1,2\}
,
\end{equation}
where $\gamma(t)$ is the time-dependent part of the interaction. The \textit{form} of $\gamma(t)$ is alternated between Eq.~(\ref{eq:gauss_i}) and Eq.~(\ref{eq:sin_i}) while performing the heuristic search. This procedure results in an interaction $\gamma(t)$ described by 3 Gaussians. The use of these 3 Gaussians for each interaction, and a suitable set of parameters $\{t_i,a_i,b_i\}$ (see Eq.~(\ref{eq:gauss_i})), produces the dynamics shown in Fig.~\ref{fig:dynamics_4q} and Fig.~\ref{fig:dynamics_4q_sbi}. We observe the system's dynamics, in Fig.~\ref{fig:dynamics_4q_sbi}(a), as a consequence of the Gaussian interaction opening the pathway for the system-effective bath population transfer every time a Gaussian pulse reaches maximum intensity. As seen in Fig.~\ref{fig:dynamics_4q_sbi}(a) and Fig.~\ref{fig:dynamics_4q_sbi}(b), one needs to coordinate the transfer of population between each subsystem, $S1$ and $S2$, and their respective effective environments, $B1$ and $B2$. The coordinated transfer can be seen in Fig.~\ref{fig:dynamics_4q_sbi}(b), where around $t=225$~a.u. the population of $B2$ transfers back to $S2$ and, to maintain the general decay behavior, a transfer of population from $S1$ to $B1$ occurs. The inverse situation can be seen at $t=300$ a.u. Lastly, it is worth noting how this process of heuristically searching for the \textit{right} interaction resembles what one does when working with coherent control of chemical reaction~\cite{Brumer1989}.

\subsection{Simulating OQS in quantum computing}
\label{sec:sqc}
\begin{figure}[!ht]
    \centering
    \includegraphics[width=1\textwidth]{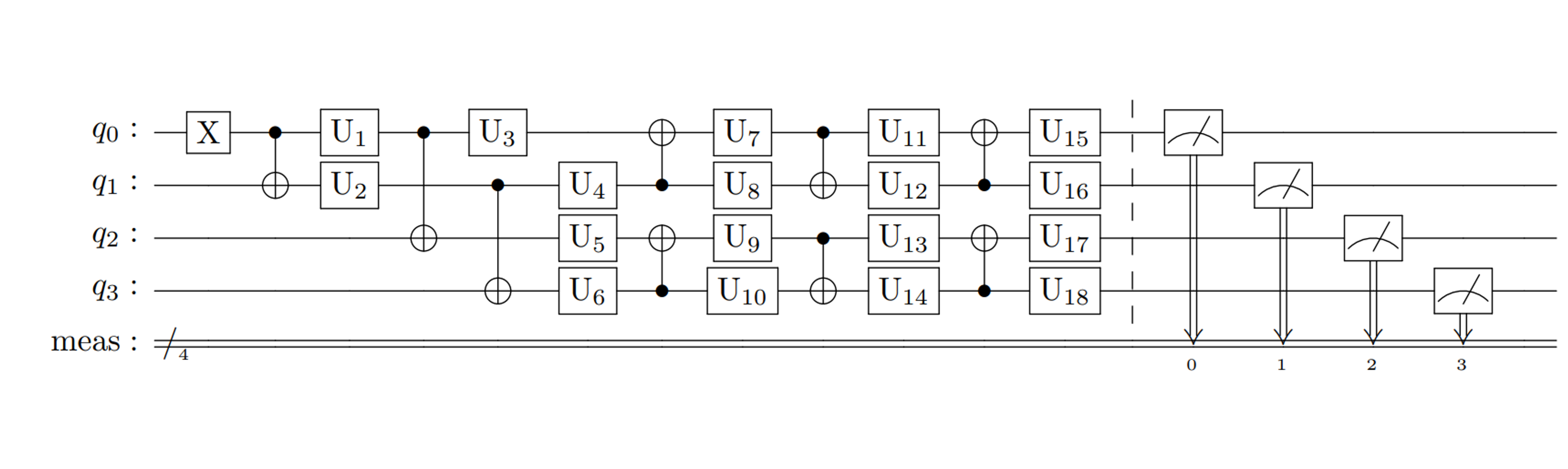}
     \caption{General circuit used to perform dynamics.}
    \label{fig:circuit_4q}
\end{figure}

All simulations in the previous sections are performed in a classical computing framework. Here, we simulate the dynamics obtained in the network of two-level systems (see Fig.~\ref{fig:dynamics_4q}) on a quantum computing framework by using Qiskit's Fake backend~\cite{qiskit2024}. Since we know the total Hamiltonian $\hat{H}_T(t)$, we can directly calculate the corresponding unitary
\begin{equation}
\label{eq:unitary_h}
\frac{d}{dt}\hat{U}(t)=-i\hat{H}_T(t)\hat{U}(t).
\end{equation}
We use Eq.~(\ref{eq:unitary_h}) as well as the fact that one can represent a general 4 qubit unitary transformation by a circuit composed of $9$ CNOT gates with $2$ general one-qubit rotations for each CNOT gate~\cite{plesch2011quantum}. This design results in the general circuit shown in Fig.~\ref{fig:circuit_4q}, in which each unitary in the set $\{U_i\}_i$ is defined by $3$ parameters. To reproduce the dynamics shown in Fig.~\ref{fig:dynamics_4q}, we calculate a unitary operator for every $4.04$ a.u. of time, leading to a total of $99$ unitaries. Each of the rotation parameters is determined using the Scipy L-BFGS-B optimization method~\cite{2020SciPy-NMeth}.

\begin{figure}[!ht]
    \centering
    \includegraphics[width=1\textwidth]{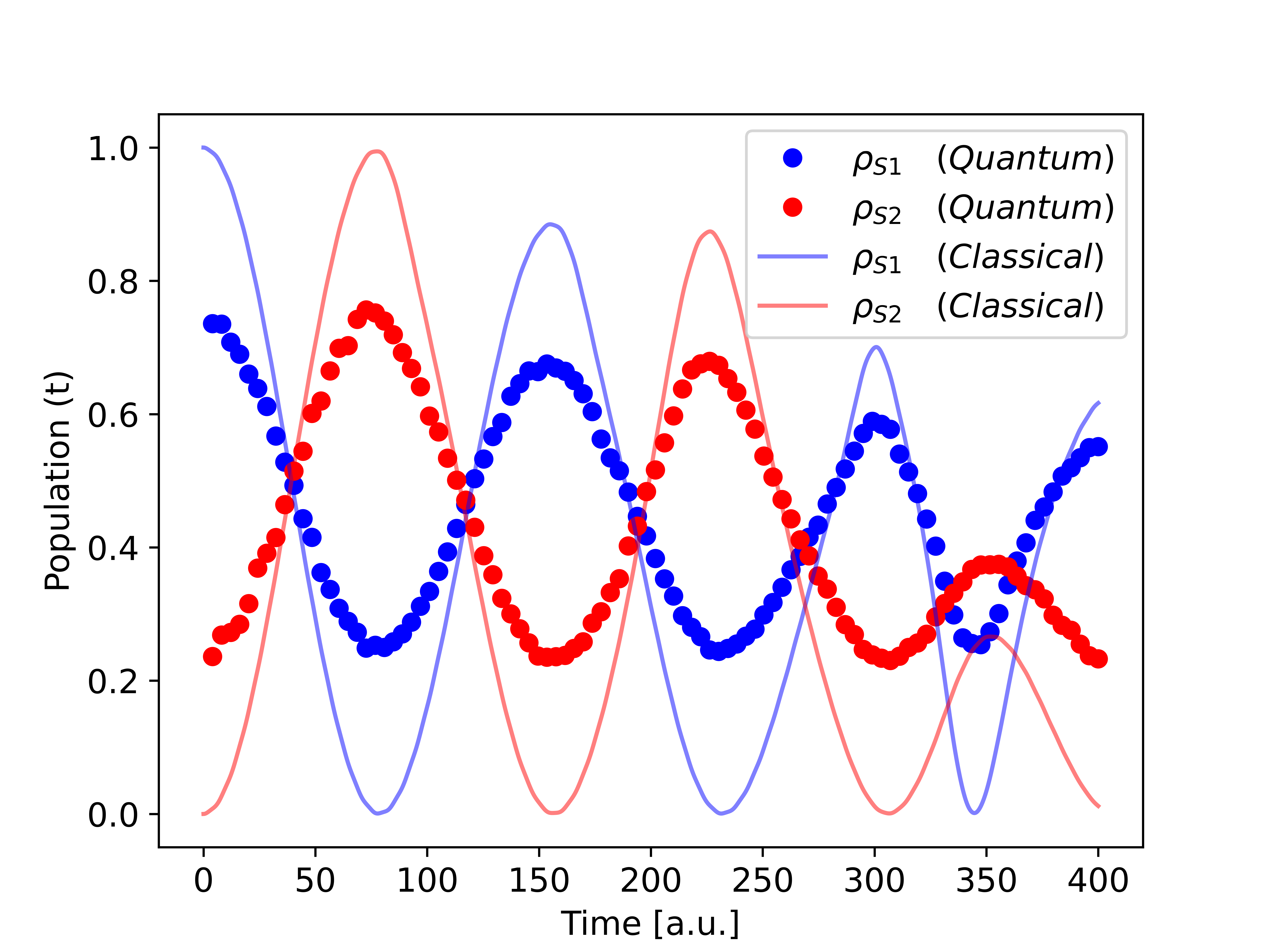}
     \caption{Dynamics of the excited states for $S1$ and $S2$\label{fig:4qubits_sim}.}
\end{figure}

By implementing this circuit in Qiskit's backend FakeBogotaV2~\cite{qiskit2024}, we were able to describe the dynamics of the network of two-level systems, as seen in Fig.~\ref{fig:4qubits_sim}, which presents the population dynamics of the excited states for both sub-systems. Ideally, the dynamics should be the same, but due to factors like decoherence, the amplitude of the population's evolution is decreased, although the general decay behavior, as well as non-Markovian behavior, is conserved. This result showcases the advantage that the purification approach has when studying OQS in a quantum device because we obtain a unitary representation of the dynamics without using classical storage or a successive dilation of the Hilbert space, which would result in the addition of ancilla qubits to the circuit when performing the dynamics (eg., Sz.-Nagy dilation).

\section{Conclusions}

In this work, we have established the main difference between the conventional approach to OQS and the density-matrix purification method, which is the non-linearity of the extension map $E_v$ for the purification method. The non-linearity in embedding the system into the environment allows access to a bigger range of dynamics, including the possibility of modeling an initially correlated system and environment. This wider range of dynamics is shown in section~\ref{sec:bta}., where we present Non-CP dynamics in Fig.~\ref{fig:dynamics_ncp}), Markovian in Fig.~\ref{fig:dynamics_2q}), and non-Markovian dynamics in Fig.~\ref{fig:dynamics_4q}). In addition, in section~\ref{sec:ca}, by following the definition of \textit{purifying} a system, which implies creating a state entangled with its effective bath, we treat an initial state that has non-trivial entanglement between the system and the environment.

Additionally, we propose a systematic way of applying the density-matrix purification approach to model Markovian and non-Markovian dynamics in Sections ~\ref{sec:bta}. 2 and 3 by relying on a heuristic search to build the system-effective bath interaction, similar to what is accomplished in the coherent control of chemical reactions. This procedure leads to the modeling of a decay process on a two-level system network, which is reminiscent of the energy transfer process in light-harvesting systems like the FMO complex. This systematic approach lays the foundations for further application and development of the purification approach. Lastly, in section~\ref{sec:sqc}, we show the application of the purification approach in a quantum computing framework by modeling the non-Markovian dynamics obtained in Section ~\ref{sec:bta}. 3, demonstrating how one can model OQS without relying on classical storage or additional ancilla qubits.

\begin{acknowledgments}
D.A.M. gratefully acknowledges support from the Department of Energy, Office of Basic Energy Sciences, Grant No. DE-SC0019215, the National Science Foundation (NSF) Grant Nos. CHE-2155082 and DMR-2037783, and the NSF QuBBE Quantum Leap Challenge Institute (NSF OMA-2121044).  This material is also based upon work supported by the U.S. Department of Energy, Office of Science, Office of Advanced Scientific Computing Research, Department of Energy Computational Science Graduate Fellowship under Award Number DE-SC0024386.  This research used resources of the National Energy Research Scientific Computing Center (NERSC), a Department of Energy Office of Science User Facility using NERSC award DDR-ERCAP0026889.
\end{acknowledgments}

\section*{Disclaimer}
This report was prepared as an account of work sponsored by an agency of the United States Government. Neither the United States Government nor any agency thereof, nor any of their employees, makes any warranty, express or implied, or assumes any legal liability or responsibility for the accuracy, completeness, or usefulness of any information, apparatus, product, or process disclosed, or represents that its use would not infringe privately owned rights. Reference herein to any specific commercial product, process, or service by trade name, trademark, manufacturer, or otherwise does not necessarily constitute or imply its endorsement, recommendation, or favoring by the United States Government or any agency thereof. The views and opinions of authors expressed herein do not necessarily state or reflect those of the United States Government or any agency thereof.


%
%

%


\bibliography{ref}

\begin{thebibliography}{95}%
\makeatletter
\providecommand \@ifxundefined [1]{%
 \@ifx{#1\undefined}
}%
\providecommand \@ifnum [1]{%
 \ifnum #1\expandafter \@firstoftwo
 \else \expandafter \@secondoftwo
 \fi
}%
\providecommand \@ifx [1]{%
 \ifx #1\expandafter \@firstoftwo
 \else \expandafter \@secondoftwo
 \fi
}%
\providecommand \natexlab [1]{#1}%
\providecommand \enquote  [1]{``#1''}%
\providecommand \bibnamefont  [1]{#1}%
\providecommand \bibfnamefont [1]{#1}%
\providecommand \citenamefont [1]{#1}%
\providecommand \href@noop [0]{\@secondoftwo}%
\providecommand \href [0]{\begingroup \@sanitize@url \@href}%
\providecommand \@href[1]{\@@startlink{#1}\@@href}%
\providecommand \@@href[1]{\endgroup#1\@@endlink}%
\providecommand \@sanitize@url [0]{\catcode `\\12\catcode `\$12\catcode `\&12\catcode `\#12\catcode `\^12\catcode `\_12\catcode `\%12\relax}%
\providecommand \@@startlink[1]{}%
\providecommand \@@endlink[0]{}%
\providecommand \url  [0]{\begingroup\@sanitize@url \@url }%
\providecommand \@url [1]{\endgroup\@href {#1}{\urlprefix }}%
\providecommand \urlprefix  [0]{URL }%
\providecommand \Eprint [0]{\href }%
\providecommand \doibase [0]{https://doi.org/}%
\providecommand \selectlanguage [0]{\@gobble}%
\providecommand \bibinfo  [0]{\@secondoftwo}%
\providecommand \bibfield  [0]{\@secondoftwo}%
\providecommand \translation [1]{[#1]}%
\providecommand \BibitemOpen [0]{}%
\providecommand \bibitemStop [0]{}%
\providecommand \bibitemNoStop [0]{.\EOS\space}%
\providecommand \EOS [0]{\spacefactor3000\relax}%
\providecommand \BibitemShut  [1]{\csname bibitem#1\endcsname}%
\let\auto@bib@innerbib\@empty
\bibitem [{\citenamefont {Mohseni}\ \emph {et~al.}(2014)\citenamefont {Mohseni}, \citenamefont {Omar}, \citenamefont {Engel},\ and\ \citenamefont {Plenio}}]{mohseni2014quantum}%
  \BibitemOpen
  \bibfield  {author} {\bibinfo {author} {\bibfnamefont {M.}~\bibnamefont {Mohseni}}, \bibinfo {author} {\bibfnamefont {Y.}~\bibnamefont {Omar}}, \bibinfo {author} {\bibfnamefont {G.~S.}\ \bibnamefont {Engel}},\ and\ \bibinfo {author} {\bibfnamefont {M.~B.}\ \bibnamefont {Plenio}},\ }\href@noop {} {\emph {\bibinfo {title} {Quantum Effects in Biology}}}\ (\bibinfo  {publisher} {Cambridge University Press},\ \bibinfo {year} {2014})\BibitemShut {NoStop}%
\bibitem [{\citenamefont {Head-Marsden}\ \emph {et~al.}(2021{\natexlab{a}})\citenamefont {Head-Marsden}, \citenamefont {Flick}, \citenamefont {Ciccarino},\ and\ \citenamefont {Narang}}]{doi:10.1021/acs.chemrev.0c00620}%
  \BibitemOpen
  \bibfield  {author} {\bibinfo {author} {\bibfnamefont {K.}~\bibnamefont {Head-Marsden}}, \bibinfo {author} {\bibfnamefont {J.}~\bibnamefont {Flick}}, \bibinfo {author} {\bibfnamefont {C.~J.}\ \bibnamefont {Ciccarino}},\ and\ \bibinfo {author} {\bibfnamefont {P.}~\bibnamefont {Narang}},\ }\bibfield  {title} {\enquote {\bibinfo {title} {Quantum information and algorithms for correlated quantum matter},}\ }\href {https://doi.org/10.1021/acs.chemrev.0c00620} {\bibfield  {journal} {\bibinfo  {journal} {Chem. Rev.}\ }\textbf {\bibinfo {volume} {121}},\ \bibinfo {pages} {3061--3120} (\bibinfo {year} {2021}{\natexlab{a}})},\ \bibinfo {note} {pMID: 33326218}\BibitemShut {NoStop}%
\bibitem [{\citenamefont {Breuer}\ and\ \citenamefont {Petruccione}(2002)}]{Breuer2002}%
  \BibitemOpen
  \bibfield  {author} {\bibinfo {author} {\bibfnamefont {H.-P.}\ \bibnamefont {Breuer}}\ and\ \bibinfo {author} {\bibfnamefont {F.}~\bibnamefont {Petruccione}},\ }\href@noop {} {\emph {\bibinfo {title} {The Theory of Open Quantum Systems}}}\ (\bibinfo  {publisher} {Oxford University Press},\ \bibinfo {year} {2002})\BibitemShut {NoStop}%
\bibitem [{\citenamefont {Jagadish}\ and\ \citenamefont {Petruccione}(2018)}]{jagadish2019}%
  \BibitemOpen
  \bibfield  {author} {\bibinfo {author} {\bibfnamefont {V.}~\bibnamefont {Jagadish}}\ and\ \bibinfo {author} {\bibfnamefont {F.}~\bibnamefont {Petruccione}},\ }\bibfield  {title} {\enquote {\bibinfo {title} {An invitation to quantum channels},}\ }\href {https://doi.org/10.12743/quanta.v7i1.77} {\bibfield  {journal} {\bibinfo  {journal} {Quanta}\ }\textbf {\bibinfo {volume} {7}},\ \bibinfo {pages} {54--67} (\bibinfo {year} {2018})}\BibitemShut {NoStop}%
\bibitem [{\citenamefont {Manzano}(2020)}]{manzano2020}%
  \BibitemOpen
  \bibfield  {author} {\bibinfo {author} {\bibfnamefont {D.}~\bibnamefont {Manzano}},\ }\bibfield  {title} {\enquote {\bibinfo {title} {{A short introduction to the Lindblad master equation}},}\ }\href {https://doi.org/10.1063/1.5115323} {\bibfield  {journal} {\bibinfo  {journal} {AIP Adv.}\ }\textbf {\bibinfo {volume} {10}},\ \bibinfo {pages} {025106} (\bibinfo {year} {2020})}\BibitemShut {NoStop}%
\bibitem [{\citenamefont {Breuer}\ \emph {et~al.}(2016)\citenamefont {Breuer}, \citenamefont {Laine}, \citenamefont {Piilo},\ and\ \citenamefont {Vacchini}}]{breuer2016}%
  \BibitemOpen
  \bibfield  {author} {\bibinfo {author} {\bibfnamefont {H.-P.}\ \bibnamefont {Breuer}}, \bibinfo {author} {\bibfnamefont {E.-M.}\ \bibnamefont {Laine}}, \bibinfo {author} {\bibfnamefont {J.}~\bibnamefont {Piilo}},\ and\ \bibinfo {author} {\bibfnamefont {B.}~\bibnamefont {Vacchini}},\ }\bibfield  {title} {\enquote {\bibinfo {title} {Colloquium: Non-markovian dynamics in open quantum systems},}\ }\href {https://doi.org/10.1103/RevModPhys.88.021002} {\bibfield  {journal} {\bibinfo  {journal} {Rev. Mod. Phys.}\ }\textbf {\bibinfo {volume} {88}},\ \bibinfo {pages} {021002} (\bibinfo {year} {2016})}\BibitemShut {NoStop}%
\bibitem [{\citenamefont {Tanimura}(2020)}]{10.1063/5.0011599}%
  \BibitemOpen
  \bibfield  {author} {\bibinfo {author} {\bibfnamefont {Y.}~\bibnamefont {Tanimura}},\ }\bibfield  {title} {\enquote {\bibinfo {title} {{Numerically “exact” approach to open quantum dynamics: The hierarchical equations of motion (HEOM)}},}\ }\href {https://doi.org/10.1063/5.0011599} {\bibfield  {journal} {\bibinfo  {journal} {J. Chem. Phys.}\ }\textbf {\bibinfo {volume} {153}},\ \bibinfo {pages} {020901} (\bibinfo {year} {2020})}\BibitemShut {NoStop}%
\bibitem [{\citenamefont {Ko}, \citenamefont {Cook},\ and\ \citenamefont {Whaley}(2022)}]{10.1063/5.0082822}%
  \BibitemOpen
  \bibfield  {author} {\bibinfo {author} {\bibfnamefont {L.}~\bibnamefont {Ko}}, \bibinfo {author} {\bibfnamefont {R.~L.}\ \bibnamefont {Cook}},\ and\ \bibinfo {author} {\bibfnamefont {K.~B.}\ \bibnamefont {Whaley}},\ }\bibfield  {title} {\enquote {\bibinfo {title} {{Dynamics of photosynthetic light harvesting systems interacting with N-photon Fock states}},}\ }\href {https://doi.org/10.1063/5.0082822} {\bibfield  {journal} {\bibinfo  {journal} {J. Chem. Phys.}\ }\textbf {\bibinfo {volume} {156}},\ \bibinfo {pages} {244108} (\bibinfo {year} {2022})}\BibitemShut {NoStop}%
\bibitem [{\citenamefont {Chan}\ \emph {et~al.}(2018)\citenamefont {Chan}, \citenamefont {Gamel}, \citenamefont {Fleming},\ and\ \citenamefont {Whaley}}]{Chan_2018}%
  \BibitemOpen
  \bibfield  {author} {\bibinfo {author} {\bibfnamefont {H.~C.~H.}\ \bibnamefont {Chan}}, \bibinfo {author} {\bibfnamefont {O.~E.}\ \bibnamefont {Gamel}}, \bibinfo {author} {\bibfnamefont {G.~R.}\ \bibnamefont {Fleming}},\ and\ \bibinfo {author} {\bibfnamefont {K.~B.}\ \bibnamefont {Whaley}},\ }\bibfield  {title} {\enquote {\bibinfo {title} {Single-photon absorption by single photosynthetic light-harvesting complexes},}\ }\href {https://doi.org/10.1088/1361-6455/aa9c95} {\bibfield  {journal} {\bibinfo  {journal} {Journal of Physics B: Atomic, Molecular and Optical Physics}\ }\textbf {\bibinfo {volume} {51}},\ \bibinfo {pages} {054002} (\bibinfo {year} {2018})}\BibitemShut {NoStop}%
\bibitem [{\citenamefont {Strathearn}\ \emph {et~al.}(2018)\citenamefont {Strathearn}, \citenamefont {Kirton}, \citenamefont {Kilda}, \citenamefont {Keeling},\ and\ \citenamefont {Lovett}}]{strathearn2018efficient}%
  \BibitemOpen
  \bibfield  {author} {\bibinfo {author} {\bibfnamefont {A.}~\bibnamefont {Strathearn}}, \bibinfo {author} {\bibfnamefont {P.}~\bibnamefont {Kirton}}, \bibinfo {author} {\bibfnamefont {D.}~\bibnamefont {Kilda}}, \bibinfo {author} {\bibfnamefont {J.}~\bibnamefont {Keeling}},\ and\ \bibinfo {author} {\bibfnamefont {B.~W.}\ \bibnamefont {Lovett}},\ }\bibfield  {title} {\enquote {\bibinfo {title} {Efficient non-markovian quantum dynamics using time-evolving matrix product operators},}\ }\href@noop {} {\bibfield  {journal} {\bibinfo  {journal} {Nat. Commun.}\ }\textbf {\bibinfo {volume} {9}},\ \bibinfo {pages} {3322} (\bibinfo {year} {2018})}\BibitemShut {NoStop}%
\bibitem [{\citenamefont {Finsterhölzl}\ \emph {et~al.}(2020)\citenamefont {Finsterhölzl}, \citenamefont {Katzer}, \citenamefont {Knorr},\ and\ \citenamefont {Carmele}}]{e22090984}%
  \BibitemOpen
  \bibfield  {author} {\bibinfo {author} {\bibfnamefont {R.}~\bibnamefont {Finsterhölzl}}, \bibinfo {author} {\bibfnamefont {M.}~\bibnamefont {Katzer}}, \bibinfo {author} {\bibfnamefont {A.}~\bibnamefont {Knorr}},\ and\ \bibinfo {author} {\bibfnamefont {A.}~\bibnamefont {Carmele}},\ }\bibfield  {title} {\enquote {\bibinfo {title} {Using matrix-product states for open quantum many-body systems: Efficient algorithms for markovian and non-markovian time-evolution},}\ }\href {https://doi.org/10.3390/e22090984} {\bibfield  {journal} {\bibinfo  {journal} {Entropy}\ }\textbf {\bibinfo {volume} {22}} (\bibinfo {year} {2020}),\ 10.3390/e22090984}\BibitemShut {NoStop}%
\bibitem [{\citenamefont {Lyu}\ \emph {et~al.}(2023)\citenamefont {Lyu}, \citenamefont {Mulvihill}, \citenamefont {Soley}, \citenamefont {Geva},\ and\ \citenamefont {Batista}}]{Lyu.2023}%
  \BibitemOpen
  \bibfield  {author} {\bibinfo {author} {\bibfnamefont {N.}~\bibnamefont {Lyu}}, \bibinfo {author} {\bibfnamefont {E.}~\bibnamefont {Mulvihill}}, \bibinfo {author} {\bibfnamefont {M.~B.}\ \bibnamefont {Soley}}, \bibinfo {author} {\bibfnamefont {E.}~\bibnamefont {Geva}},\ and\ \bibinfo {author} {\bibfnamefont {V.~S.}\ \bibnamefont {Batista}},\ }\bibfield  {title} {\enquote {\bibinfo {title} {{Tensor-Train Thermo-Field Memory Kernels for Generalized Quantum Master Equations}},}\ }\href {https://doi.org/10.1021/acs.jctc.2c00892} {\bibfield  {journal} {\bibinfo  {journal} {Journal of Chemical Theory and Computation}\ }\textbf {\bibinfo {volume} {19}},\ \bibinfo {pages} {1111--1129} (\bibinfo {year} {2023})},\ \Eprint {https://arxiv.org/abs/2208.14273} {2208.14273} \BibitemShut {NoStop}%
\bibitem [{\citenamefont {Clark}\ \emph {et~al.}(2010)\citenamefont {Clark}, \citenamefont {Prior}, \citenamefont {Hartmann}, \citenamefont {Jaksch},\ and\ \citenamefont {Plenio}}]{Clark_2010}%
  \BibitemOpen
  \bibfield  {author} {\bibinfo {author} {\bibfnamefont {S.~R.}\ \bibnamefont {Clark}}, \bibinfo {author} {\bibfnamefont {J.}~\bibnamefont {Prior}}, \bibinfo {author} {\bibfnamefont {M.~J.}\ \bibnamefont {Hartmann}}, \bibinfo {author} {\bibfnamefont {D.}~\bibnamefont {Jaksch}},\ and\ \bibinfo {author} {\bibfnamefont {M.~B.}\ \bibnamefont {Plenio}},\ }\bibfield  {title} {\enquote {\bibinfo {title} {Exact matrix product solutions in the heisenberg picture of an open quantum spin chain},}\ }\href {https://doi.org/10.1088/1367-2630/12/2/025005} {\bibfield  {journal} {\bibinfo  {journal} {New J. Phys.}\ }\textbf {\bibinfo {volume} {12}},\ \bibinfo {pages} {025005} (\bibinfo {year} {2010})}\BibitemShut {NoStop}%
\bibitem [{\citenamefont {Jaschke}, \citenamefont {Wall},\ and\ \citenamefont {Carr}(2018)}]{JASCHKE201859}%
  \BibitemOpen
  \bibfield  {author} {\bibinfo {author} {\bibfnamefont {D.}~\bibnamefont {Jaschke}}, \bibinfo {author} {\bibfnamefont {M.~L.}\ \bibnamefont {Wall}},\ and\ \bibinfo {author} {\bibfnamefont {L.~D.}\ \bibnamefont {Carr}},\ }\bibfield  {title} {\enquote {\bibinfo {title} {Open source matrix product states: Opening ways to simulate entangled many-body quantum systems in one dimension},}\ }\href {https://doi.org/https://doi.org/10.1016/j.cpc.2017.12.015} {\bibfield  {journal} {\bibinfo  {journal} {Comput. Phys. Commun.}\ }\textbf {\bibinfo {volume} {225}},\ \bibinfo {pages} {59--91} (\bibinfo {year} {2018})}\BibitemShut {NoStop}%
\bibitem [{\citenamefont {Vicentini}\ \emph {et~al.}(2019)\citenamefont {Vicentini}, \citenamefont {Biella}, \citenamefont {Regnault},\ and\ \citenamefont {Ciuti}}]{vicentini2019variational}%
  \BibitemOpen
  \bibfield  {author} {\bibinfo {author} {\bibfnamefont {F.}~\bibnamefont {Vicentini}}, \bibinfo {author} {\bibfnamefont {A.}~\bibnamefont {Biella}}, \bibinfo {author} {\bibfnamefont {N.}~\bibnamefont {Regnault}},\ and\ \bibinfo {author} {\bibfnamefont {C.}~\bibnamefont {Ciuti}},\ }\bibfield  {title} {\enquote {\bibinfo {title} {Variational neural-network ansatz for steady states in open quantum systems},}\ }\href {https://doi.org/10.1103/PhysRevLett.122.250503} {\bibfield  {journal} {\bibinfo  {journal} {Phys. Rev. Lett.}\ }\textbf {\bibinfo {volume} {122}},\ \bibinfo {pages} {250503} (\bibinfo {year} {2019})}\BibitemShut {NoStop}%
\bibitem [{\citenamefont {Yoshioka}\ and\ \citenamefont {Hamazaki}(2019)}]{PhysRevB.99.214306}%
  \BibitemOpen
  \bibfield  {author} {\bibinfo {author} {\bibfnamefont {N.}~\bibnamefont {Yoshioka}}\ and\ \bibinfo {author} {\bibfnamefont {R.}~\bibnamefont {Hamazaki}},\ }\bibfield  {title} {\enquote {\bibinfo {title} {Constructing neural stationary states for open quantum many-body systems},}\ }\href {https://doi.org/10.1103/PhysRevB.99.214306} {\bibfield  {journal} {\bibinfo  {journal} {Phys. Rev. B}\ }\textbf {\bibinfo {volume} {99}},\ \bibinfo {pages} {214306} (\bibinfo {year} {2019})}\BibitemShut {NoStop}%
\bibitem [{\citenamefont {Hartmann}\ and\ \citenamefont {Carleo}(2019)}]{PhysRevLett.122.250502}%
  \BibitemOpen
  \bibfield  {author} {\bibinfo {author} {\bibfnamefont {M.~J.}\ \bibnamefont {Hartmann}}\ and\ \bibinfo {author} {\bibfnamefont {G.}~\bibnamefont {Carleo}},\ }\bibfield  {title} {\enquote {\bibinfo {title} {Neural-network approach to dissipative quantum many-body dynamics},}\ }\href {https://doi.org/10.1103/PhysRevLett.122.250502} {\bibfield  {journal} {\bibinfo  {journal} {Phys. Rev. Lett.}\ }\textbf {\bibinfo {volume} {122}},\ \bibinfo {pages} {250502} (\bibinfo {year} {2019})}\BibitemShut {NoStop}%
\bibitem [{\citenamefont {Carleo}\ and\ \citenamefont {Troyer}(2017)}]{doi:10.1126/science.aag2302}%
  \BibitemOpen
  \bibfield  {author} {\bibinfo {author} {\bibfnamefont {G.}~\bibnamefont {Carleo}}\ and\ \bibinfo {author} {\bibfnamefont {M.}~\bibnamefont {Troyer}},\ }\bibfield  {title} {\enquote {\bibinfo {title} {Solving the quantum many-body problem with artificial neural networks},}\ }\href {https://doi.org/10.1126/science.aag2302} {\bibfield  {journal} {\bibinfo  {journal} {Science}\ }\textbf {\bibinfo {volume} {355}},\ \bibinfo {pages} {602--606} (\bibinfo {year} {2017})}\BibitemShut {NoStop}%
\bibitem [{\citenamefont {Bacon}\ \emph {et~al.}(2001)\citenamefont {Bacon}, \citenamefont {Childs}, \citenamefont {Chuang}, \citenamefont {Kempe}, \citenamefont {Leung},\ and\ \citenamefont {Zhou}}]{Bacon.2001}%
  \BibitemOpen
  \bibfield  {author} {\bibinfo {author} {\bibfnamefont {D.}~\bibnamefont {Bacon}}, \bibinfo {author} {\bibfnamefont {A.~M.}\ \bibnamefont {Childs}}, \bibinfo {author} {\bibfnamefont {I.~L.}\ \bibnamefont {Chuang}}, \bibinfo {author} {\bibfnamefont {J.}~\bibnamefont {Kempe}}, \bibinfo {author} {\bibfnamefont {D.~W.}\ \bibnamefont {Leung}},\ and\ \bibinfo {author} {\bibfnamefont {X.}~\bibnamefont {Zhou}},\ }\bibfield  {title} {\enquote {\bibinfo {title} {{Universal simulation of Markovian quantum dynamics}},}\ }\href {https://doi.org/10.1103/physreva.64.062302} {\bibfield  {journal} {\bibinfo  {journal} {Phys. Rev. A}\ }\textbf {\bibinfo {volume} {64}},\ \bibinfo {pages} {062302} (\bibinfo {year} {2001})},\ \Eprint {https://arxiv.org/abs/quant-ph/0008070} {quant-ph/0008070} \BibitemShut {NoStop}%
\bibitem [{\citenamefont {Müller}\ \emph {et~al.}(2011)\citenamefont {Müller}, \citenamefont {Hammerer}, \citenamefont {Zhou}, \citenamefont {Roos},\ and\ \citenamefont {Zoller}}]{Müller.2011}%
  \BibitemOpen
  \bibfield  {author} {\bibinfo {author} {\bibfnamefont {M.}~\bibnamefont {Müller}}, \bibinfo {author} {\bibfnamefont {K.}~\bibnamefont {Hammerer}}, \bibinfo {author} {\bibfnamefont {Y.~L.}\ \bibnamefont {Zhou}}, \bibinfo {author} {\bibfnamefont {C.~F.}\ \bibnamefont {Roos}},\ and\ \bibinfo {author} {\bibfnamefont {P.}~\bibnamefont {Zoller}},\ }\bibfield  {title} {\enquote {\bibinfo {title} {{Simulating open quantum systems: from many-body interactions to stabilizer pumping}},}\ }\href {https://doi.org/10.1088/1367-2630/13/8/085007} {\bibfield  {journal} {\bibinfo  {journal} {New J. Phys.}\ }\textbf {\bibinfo {volume} {13}},\ \bibinfo {pages} {085007} (\bibinfo {year} {2011})},\ \Eprint {https://arxiv.org/abs/1104.2507} {1104.2507} \BibitemShut {NoStop}%
\bibitem [{\citenamefont {Wang}, \citenamefont {Ashhab},\ and\ \citenamefont {Nori}(2011)}]{Wang.2011}%
  \BibitemOpen
  \bibfield  {author} {\bibinfo {author} {\bibfnamefont {H.}~\bibnamefont {Wang}}, \bibinfo {author} {\bibfnamefont {S.}~\bibnamefont {Ashhab}},\ and\ \bibinfo {author} {\bibfnamefont {F.}~\bibnamefont {Nori}},\ }\bibfield  {title} {\enquote {\bibinfo {title} {{Quantum algorithm for simulating the dynamics of an open quantum system}},}\ }\href {https://doi.org/10.1103/physreva.83.062317} {\bibfield  {journal} {\bibinfo  {journal} {Phys. Rev. A}\ }\textbf {\bibinfo {volume} {83}},\ \bibinfo {pages} {062317} (\bibinfo {year} {2011})},\ \Eprint {https://arxiv.org/abs/1103.3377} {1103.3377} \BibitemShut {NoStop}%
\bibitem [{\citenamefont {Mostame}\ \emph {et~al.}(2012)\citenamefont {Mostame}, \citenamefont {Rebentrost}, \citenamefont {Eisfeld}, \citenamefont {Kerman}, \citenamefont {Tsomokos},\ and\ \citenamefont {Aspuru-Guzik}}]{Mostame.2012}%
  \BibitemOpen
  \bibfield  {author} {\bibinfo {author} {\bibfnamefont {S.}~\bibnamefont {Mostame}}, \bibinfo {author} {\bibfnamefont {P.}~\bibnamefont {Rebentrost}}, \bibinfo {author} {\bibfnamefont {A.}~\bibnamefont {Eisfeld}}, \bibinfo {author} {\bibfnamefont {A.~J.}\ \bibnamefont {Kerman}}, \bibinfo {author} {\bibfnamefont {D.~I.}\ \bibnamefont {Tsomokos}},\ and\ \bibinfo {author} {\bibfnamefont {A.}~\bibnamefont {Aspuru-Guzik}},\ }\bibfield  {title} {\enquote {\bibinfo {title} {{Quantum simulator of an open quantum system using superconducting qubits: exciton transport in photosynthetic complexes}},}\ }\href {https://doi.org/10.1088/1367-2630/14/10/105013} {\bibfield  {journal} {\bibinfo  {journal} {New J. Phys.}\ }\textbf {\bibinfo {volume} {14}},\ \bibinfo {pages} {105013} (\bibinfo {year} {2012})},\ \Eprint {https://arxiv.org/abs/1106.1683} {1106.1683} \BibitemShut {NoStop}%
\bibitem [{\citenamefont {Sweke}\ \emph {et~al.}(2015)\citenamefont {Sweke}, \citenamefont {Sinayskiy}, \citenamefont {Bernard},\ and\ \citenamefont {Petruccione}}]{Sweke.2015}%
  \BibitemOpen
  \bibfield  {author} {\bibinfo {author} {\bibfnamefont {R.}~\bibnamefont {Sweke}}, \bibinfo {author} {\bibfnamefont {I.}~\bibnamefont {Sinayskiy}}, \bibinfo {author} {\bibfnamefont {D.}~\bibnamefont {Bernard}},\ and\ \bibinfo {author} {\bibfnamefont {F.}~\bibnamefont {Petruccione}},\ }\bibfield  {title} {\enquote {\bibinfo {title} {{Universal simulation of Markovian open quantum systems}},}\ }\href {https://doi.org/10.1103/physreva.91.062308} {\bibfield  {journal} {\bibinfo  {journal} {Phys. Rev. A}\ }\textbf {\bibinfo {volume} {91}},\ \bibinfo {pages} {062308} (\bibinfo {year} {2015})},\ \Eprint {https://arxiv.org/abs/1503.05028} {1503.05028} \BibitemShut {NoStop}%
\bibitem [{\citenamefont {Mostame}\ \emph {et~al.}(2016)\citenamefont {Mostame}, \citenamefont {Huh}, \citenamefont {Kreisbeck}, \citenamefont {Kerman}, \citenamefont {Fujita}, \citenamefont {Eisfeld},\ and\ \citenamefont {Aspuru-Guzik}}]{Mostame.2016}%
  \BibitemOpen
  \bibfield  {author} {\bibinfo {author} {\bibfnamefont {S.}~\bibnamefont {Mostame}}, \bibinfo {author} {\bibfnamefont {J.}~\bibnamefont {Huh}}, \bibinfo {author} {\bibfnamefont {C.}~\bibnamefont {Kreisbeck}}, \bibinfo {author} {\bibfnamefont {A.~J.}\ \bibnamefont {Kerman}}, \bibinfo {author} {\bibfnamefont {T.}~\bibnamefont {Fujita}}, \bibinfo {author} {\bibfnamefont {A.}~\bibnamefont {Eisfeld}},\ and\ \bibinfo {author} {\bibfnamefont {A.}~\bibnamefont {Aspuru-Guzik}},\ }\bibfield  {title} {\enquote {\bibinfo {title} {{Emulation of complex open quantum systems using superconducting qubits}},}\ }\href {https://doi.org/10.1007/s11128-016-1489-3} {\bibfield  {journal} {\bibinfo  {journal} {Quantum Inf. Process.}\ }\textbf {\bibinfo {volume} {16}},\ \bibinfo {pages} {44} (\bibinfo {year} {2016})},\ \Eprint {https://arxiv.org/abs/1502.00962} {1502.00962} \BibitemShut {NoStop}%
\bibitem [{\citenamefont {Sweke}\ \emph {et~al.}(2016)\citenamefont {Sweke}, \citenamefont {Sanz}, \citenamefont {Sinayskiy}, \citenamefont {Petruccione},\ and\ \citenamefont {Solano}}]{Sweke.2016}%
  \BibitemOpen
  \bibfield  {author} {\bibinfo {author} {\bibfnamefont {R.}~\bibnamefont {Sweke}}, \bibinfo {author} {\bibfnamefont {M.}~\bibnamefont {Sanz}}, \bibinfo {author} {\bibfnamefont {I.}~\bibnamefont {Sinayskiy}}, \bibinfo {author} {\bibfnamefont {F.}~\bibnamefont {Petruccione}},\ and\ \bibinfo {author} {\bibfnamefont {E.}~\bibnamefont {Solano}},\ }\bibfield  {title} {\enquote {\bibinfo {title} {{Digital quantum simulation of many-body non-Markovian dynamics}},}\ }\href {https://doi.org/10.1103/physreva.94.022317} {\bibfield  {journal} {\bibinfo  {journal} {Phys. Rev. A}\ }\textbf {\bibinfo {volume} {94}},\ \bibinfo {pages} {022317} (\bibinfo {year} {2016})},\ \Eprint {https://arxiv.org/abs/1604.00203} {1604.00203} \BibitemShut {NoStop}%
\bibitem [{\citenamefont {Suri}\ \emph {et~al.}(2018)\citenamefont {Suri}, \citenamefont {Binder}, \citenamefont {Muralidharan},\ and\ \citenamefont {Vinjanampathy}}]{Suri.2018}%
  \BibitemOpen
  \bibfield  {author} {\bibinfo {author} {\bibfnamefont {N.}~\bibnamefont {Suri}}, \bibinfo {author} {\bibfnamefont {F.~C.}\ \bibnamefont {Binder}}, \bibinfo {author} {\bibfnamefont {B.}~\bibnamefont {Muralidharan}},\ and\ \bibinfo {author} {\bibfnamefont {S.}~\bibnamefont {Vinjanampathy}},\ }\bibfield  {title} {\enquote {\bibinfo {title} {{Speeding up thermalisation via open quantum system variational optimisation}},}\ }\href {https://doi.org/10.1140/epjst/e2018-00125-6} {\bibfield  {journal} {\bibinfo  {journal} {Eur. Phys. J. Spec. Top.}\ }\textbf {\bibinfo {volume} {227}},\ \bibinfo {pages} {203--216} (\bibinfo {year} {2018})},\ \Eprint {https://arxiv.org/abs/1711.08776} {1711.08776} \BibitemShut {NoStop}%
\bibitem [{\citenamefont {Inoue}\ and\ \citenamefont {Fukumoto}(2018)}]{Inoue.2018}%
  \BibitemOpen
  \bibfield  {author} {\bibinfo {author} {\bibfnamefont {K.}~\bibnamefont {Inoue}}\ and\ \bibinfo {author} {\bibfnamefont {Y.}~\bibnamefont {Fukumoto}},\ }\bibfield  {title} {\enquote {\bibinfo {title} {{Typical Purification Reproducing the Time Evolution of an Open Quantum System}},}\ }\href {https://doi.org/10.48550/arxiv.1811.00235} {\bibfield  {journal} {\bibinfo  {journal} {arXiv}\ } (\bibinfo {year} {2018}),\ 10.48550/arxiv.1811.00235},\ \Eprint {https://arxiv.org/abs/1811.00235} {1811.00235} \BibitemShut {NoStop}%
\bibitem [{\citenamefont {Gupta}\ and\ \citenamefont {Chandrashekar}(2020{\natexlab{a}})}]{Gupta.2020}%
  \BibitemOpen
  \bibfield  {author} {\bibinfo {author} {\bibfnamefont {P.}~\bibnamefont {Gupta}}\ and\ \bibinfo {author} {\bibfnamefont {C.~M.}\ \bibnamefont {Chandrashekar}},\ }\bibfield  {title} {\enquote {\bibinfo {title} {{Optimal quantum simulation of open quantum systems}},}\ }\href {https://doi.org/10.48550/arxiv.2012.07540} {\bibfield  {journal} {\bibinfo  {journal} {arXiv}\ } (\bibinfo {year} {2020}{\natexlab{a}}),\ 10.48550/arxiv.2012.07540},\ \Eprint {https://arxiv.org/abs/2012.07540} {2012.07540} \BibitemShut {NoStop}%
\bibitem [{\citenamefont {García-Pérez}, \citenamefont {Rossi},\ and\ \citenamefont {Maniscalco}(2020)}]{García-Pérez.2020}%
  \BibitemOpen
  \bibfield  {author} {\bibinfo {author} {\bibfnamefont {G.}~\bibnamefont {García-Pérez}}, \bibinfo {author} {\bibfnamefont {M.~A.~C.}\ \bibnamefont {Rossi}},\ and\ \bibinfo {author} {\bibfnamefont {S.}~\bibnamefont {Maniscalco}},\ }\bibfield  {title} {\enquote {\bibinfo {title} {{IBM Q Experience as a versatile experimental testbed for simulating open quantum systems}},}\ }\href {https://doi.org/10.1038/s41534-019-0235-y} {\bibfield  {journal} {\bibinfo  {journal} {Npj Quantum Inf.}\ }\textbf {\bibinfo {volume} {6}},\ \bibinfo {pages} {1} (\bibinfo {year} {2020})},\ \Eprint {https://arxiv.org/abs/1906.07099} {1906.07099} \BibitemShut {NoStop}%
\bibitem [{\citenamefont {Endo}\ \emph {et~al.}(2020)\citenamefont {Endo}, \citenamefont {Sun}, \citenamefont {Li}, \citenamefont {Benjamin},\ and\ \citenamefont {Yuan}}]{Endo.2020}%
  \BibitemOpen
  \bibfield  {author} {\bibinfo {author} {\bibfnamefont {S.}~\bibnamefont {Endo}}, \bibinfo {author} {\bibfnamefont {J.}~\bibnamefont {Sun}}, \bibinfo {author} {\bibfnamefont {Y.}~\bibnamefont {Li}}, \bibinfo {author} {\bibfnamefont {S.~C.}\ \bibnamefont {Benjamin}},\ and\ \bibinfo {author} {\bibfnamefont {X.}~\bibnamefont {Yuan}},\ }\bibfield  {title} {\enquote {\bibinfo {title} {{Variational Quantum Simulation of General Processes}},}\ }\href {https://doi.org/10.1103/physrevlett.125.010501} {\bibfield  {journal} {\bibinfo  {journal} {Phys. Rev. Lett.}\ }\textbf {\bibinfo {volume} {125}},\ \bibinfo {pages} {010501} (\bibinfo {year} {2020})},\ \Eprint {https://arxiv.org/abs/1812.08778} {1812.08778} \BibitemShut {NoStop}%
\bibitem [{\citenamefont {Patsch}, \citenamefont {Maniscalco},\ and\ \citenamefont {Koch}(2020)}]{Patsch.2020}%
  \BibitemOpen
  \bibfield  {author} {\bibinfo {author} {\bibfnamefont {S.}~\bibnamefont {Patsch}}, \bibinfo {author} {\bibfnamefont {S.}~\bibnamefont {Maniscalco}},\ and\ \bibinfo {author} {\bibfnamefont {C.~P.}\ \bibnamefont {Koch}},\ }\bibfield  {title} {\enquote {\bibinfo {title} {{Simulation of open-quantum-system dynamics using the quantum Zeno effect}},}\ }\href {https://doi.org/10.1103/physrevresearch.2.023133} {\bibfield  {journal} {\bibinfo  {journal} {Phys. Rev. Research}\ }\textbf {\bibinfo {volume} {2}},\ \bibinfo {pages} {023133} (\bibinfo {year} {2020})},\ \Eprint {https://arxiv.org/abs/1906.11492} {1906.11492} \BibitemShut {NoStop}%
\bibitem [{\citenamefont {Gupta}\ and\ \citenamefont {Chandrashekar}(2020{\natexlab{b}})}]{Gupta.2020xew}%
  \BibitemOpen
  \bibfield  {author} {\bibinfo {author} {\bibfnamefont {P.}~\bibnamefont {Gupta}}\ and\ \bibinfo {author} {\bibfnamefont {C.~M.}\ \bibnamefont {Chandrashekar}},\ }\bibfield  {title} {\enquote {\bibinfo {title} {{Digital quantum simulation framework for energy transport in an open quantum system}},}\ }\href {https://doi.org/10.1088/1367-2630/abcdc9} {\bibfield  {journal} {\bibinfo  {journal} {New J. Phys.}\ }\textbf {\bibinfo {volume} {22}},\ \bibinfo {pages} {123027} (\bibinfo {year} {2020}{\natexlab{b}})},\ \Eprint {https://arxiv.org/abs/2006.14136} {2006.14136} \BibitemShut {NoStop}%
\bibitem [{\citenamefont {Hu}, \citenamefont {Xia},\ and\ \citenamefont {Kais}(2020)}]{Hu.2020}%
  \BibitemOpen
  \bibfield  {author} {\bibinfo {author} {\bibfnamefont {Z.}~\bibnamefont {Hu}}, \bibinfo {author} {\bibfnamefont {R.}~\bibnamefont {Xia}},\ and\ \bibinfo {author} {\bibfnamefont {S.}~\bibnamefont {Kais}},\ }\bibfield  {title} {\enquote {\bibinfo {title} {{A quantum algorithm for evolving open quantum dynamics on quantum computing devices}},}\ }\href {https://doi.org/10.1038/s41598-020-60321-x} {\bibfield  {journal} {\bibinfo  {journal} {Sci. Rep.}\ }\textbf {\bibinfo {volume} {10}},\ \bibinfo {pages} {3301} (\bibinfo {year} {2020})},\ \Eprint {https://arxiv.org/abs/1904.00910} {1904.00910} \BibitemShut {NoStop}%
\bibitem [{\citenamefont {Lin}\ \emph {et~al.}(2021)\citenamefont {Lin}, \citenamefont {Dilip}, \citenamefont {Green}, \citenamefont {Smith},\ and\ \citenamefont {Pollmann}}]{Lin.2021}%
  \BibitemOpen
  \bibfield  {author} {\bibinfo {author} {\bibfnamefont {S.-H.}\ \bibnamefont {Lin}}, \bibinfo {author} {\bibfnamefont {R.}~\bibnamefont {Dilip}}, \bibinfo {author} {\bibfnamefont {A.~G.}\ \bibnamefont {Green}}, \bibinfo {author} {\bibfnamefont {A.}~\bibnamefont {Smith}},\ and\ \bibinfo {author} {\bibfnamefont {F.}~\bibnamefont {Pollmann}},\ }\bibfield  {title} {\enquote {\bibinfo {title} {{Real- and Imaginary-Time Evolution with Compressed Quantum Circuits}},}\ }\href {https://doi.org/10.1103/prxquantum.2.010342} {\bibfield  {journal} {\bibinfo  {journal} {PRX Quantum}\ }\textbf {\bibinfo {volume} {2}} (\bibinfo {year} {2021}),\ 10.1103/prxquantum.2.010342},\ \Eprint {https://arxiv.org/abs/2008.10322} {2008.10322} \BibitemShut {NoStop}%
\bibitem [{\citenamefont {Lee}\ \emph {et~al.}(2021)\citenamefont {Lee}, \citenamefont {Patil}, \citenamefont {Zhang},\ and\ \citenamefont {Hsieh}}]{Lee.2021}%
  \BibitemOpen
  \bibfield  {author} {\bibinfo {author} {\bibfnamefont {C.~K.}\ \bibnamefont {Lee}}, \bibinfo {author} {\bibfnamefont {P.}~\bibnamefont {Patil}}, \bibinfo {author} {\bibfnamefont {S.}~\bibnamefont {Zhang}},\ and\ \bibinfo {author} {\bibfnamefont {C.~Y.}\ \bibnamefont {Hsieh}},\ }\bibfield  {title} {\enquote {\bibinfo {title} {{Neural-network variational quantum algorithm for simulating many-body dynamics}},}\ }\href {https://doi.org/10.1103/physrevresearch.3.023095} {\bibfield  {journal} {\bibinfo  {journal} {Phys. Rev. Research}\ }\textbf {\bibinfo {volume} {3}},\ \bibinfo {pages} {023095} (\bibinfo {year} {2021})},\ \Eprint {https://arxiv.org/abs/2008.13329} {2008.13329} \BibitemShut {NoStop}%
\bibitem [{\citenamefont {Schlimgen}\ \emph {et~al.}(2021)\citenamefont {Schlimgen}, \citenamefont {Head-Marsden}, \citenamefont {Sager}, \citenamefont {Narang},\ and\ \citenamefont {Mazziotti}}]{Schlimgen.2021}%
  \BibitemOpen
  \bibfield  {author} {\bibinfo {author} {\bibfnamefont {A.~W.}\ \bibnamefont {Schlimgen}}, \bibinfo {author} {\bibfnamefont {K.}~\bibnamefont {Head-Marsden}}, \bibinfo {author} {\bibfnamefont {L.~M.}\ \bibnamefont {Sager}}, \bibinfo {author} {\bibfnamefont {P.}~\bibnamefont {Narang}},\ and\ \bibinfo {author} {\bibfnamefont {D.~A.}\ \bibnamefont {Mazziotti}},\ }\bibfield  {title} {\enquote {\bibinfo {title} {{Quantum Simulation of Open Quantum Systems Using a Unitary Decomposition of Operators}},}\ }\href {https://doi.org/10.1103/physrevlett.127.270503} {\bibfield  {journal} {\bibinfo  {journal} {Phys. Rev. Lett.}\ }\textbf {\bibinfo {volume} {127}},\ \bibinfo {pages} {270503} (\bibinfo {year} {2021})},\ \Eprint {https://arxiv.org/abs/2106.12588} {2106.12588} \BibitemShut {NoStop}%
\bibitem [{\citenamefont {Head-Marsden}\ \emph {et~al.}(2021{\natexlab{b}})\citenamefont {Head-Marsden}, \citenamefont {Krastanov}, \citenamefont {Mazziotti},\ and\ \citenamefont {Narang}}]{Head-Marsden.202180m}%
  \BibitemOpen
  \bibfield  {author} {\bibinfo {author} {\bibfnamefont {K.}~\bibnamefont {Head-Marsden}}, \bibinfo {author} {\bibfnamefont {S.}~\bibnamefont {Krastanov}}, \bibinfo {author} {\bibfnamefont {D.~A.}\ \bibnamefont {Mazziotti}},\ and\ \bibinfo {author} {\bibfnamefont {P.}~\bibnamefont {Narang}},\ }\bibfield  {title} {\enquote {\bibinfo {title} {{Capturing non-Markovian dynamics on near-term quantum computers}},}\ }\href {https://doi.org/10.1103/physrevresearch.3.013182} {\bibfield  {journal} {\bibinfo  {journal} {Phys. Rev. Research}\ }\textbf {\bibinfo {volume} {3}},\ \bibinfo {pages} {013182} (\bibinfo {year} {2021}{\natexlab{b}})},\ \Eprint {https://arxiv.org/abs/2005.00029} {2005.00029} \BibitemShut {NoStop}%
\bibitem [{\citenamefont {Gaikwad}, \citenamefont {Arvind},\ and\ \citenamefont {Dorai}(2022)}]{Gaikwad.2022}%
  \BibitemOpen
  \bibfield  {author} {\bibinfo {author} {\bibfnamefont {A.}~\bibnamefont {Gaikwad}}, \bibinfo {author} {\bibnamefont {Arvind}},\ and\ \bibinfo {author} {\bibfnamefont {K.}~\bibnamefont {Dorai}},\ }\bibfield  {title} {\enquote {\bibinfo {title} {{Simulating open quantum dynamics on an NMR quantum processor using the Sz.-Nagy dilation algorithm}},}\ }\href {https://doi.org/10.1103/physreva.106.022424} {\bibfield  {journal} {\bibinfo  {journal} {Phys. Rev. A}\ }\textbf {\bibinfo {volume} {106}},\ \bibinfo {pages} {022424} (\bibinfo {year} {2022})},\ \Eprint {https://arxiv.org/abs/2201.07687} {2201.07687} \BibitemShut {NoStop}%
\bibitem [{\citenamefont {Jong}\ \emph {et~al.}(2022)\citenamefont {Jong}, \citenamefont {Lee}, \citenamefont {Mulligan}, \citenamefont {Płoskoń}, \citenamefont {Ringer},\ and\ \citenamefont {Yao}}]{Jong.2022}%
  \BibitemOpen
  \bibfield  {author} {\bibinfo {author} {\bibfnamefont {W.~A.~d.}\ \bibnamefont {Jong}}, \bibinfo {author} {\bibfnamefont {K.}~\bibnamefont {Lee}}, \bibinfo {author} {\bibfnamefont {J.}~\bibnamefont {Mulligan}}, \bibinfo {author} {\bibfnamefont {M.}~\bibnamefont {Płoskoń}}, \bibinfo {author} {\bibfnamefont {F.}~\bibnamefont {Ringer}},\ and\ \bibinfo {author} {\bibfnamefont {X.}~\bibnamefont {Yao}},\ }\bibfield  {title} {\enquote {\bibinfo {title} {{Quantum simulation of nonequilibrium dynamics and thermalization in the Schwinger model}},}\ }\href {https://doi.org/10.1103/physrevd.106.054508} {\bibfield  {journal} {\bibinfo  {journal} {Phys. Rev. D}\ }\textbf {\bibinfo {volume} {106}},\ \bibinfo {pages} {054508} (\bibinfo {year} {2022})},\ \Eprint {https://arxiv.org/abs/2106.08394} {2106.08394} \BibitemShut {NoStop}%
\bibitem [{\citenamefont {Kamakari}\ \emph {et~al.}(2022)\citenamefont {Kamakari}, \citenamefont {Sun}, \citenamefont {Motta},\ and\ \citenamefont {Minnich}}]{Kamakari.2022}%
  \BibitemOpen
  \bibfield  {author} {\bibinfo {author} {\bibfnamefont {H.}~\bibnamefont {Kamakari}}, \bibinfo {author} {\bibfnamefont {S.-N.}\ \bibnamefont {Sun}}, \bibinfo {author} {\bibfnamefont {M.}~\bibnamefont {Motta}},\ and\ \bibinfo {author} {\bibfnamefont {A.~J.}\ \bibnamefont {Minnich}},\ }\bibfield  {title} {\enquote {\bibinfo {title} {{Digital Quantum Simulation of Open Quantum Systems Using Quantum Imaginary–Time Evolution}},}\ }\href {https://doi.org/10.1103/prxquantum.3.010320} {\bibfield  {journal} {\bibinfo  {journal} {PRX Quantum}\ }\textbf {\bibinfo {volume} {3}},\ \bibinfo {pages} {010320} (\bibinfo {year} {2022})},\ \Eprint {https://arxiv.org/abs/2104.07823} {2104.07823} \BibitemShut {NoStop}%
\bibitem [{\citenamefont {Tornow}, \citenamefont {Gehrke},\ and\ \citenamefont {Helmbrecht}(2022)}]{Tornow.2022}%
  \BibitemOpen
  \bibfield  {author} {\bibinfo {author} {\bibfnamefont {S.}~\bibnamefont {Tornow}}, \bibinfo {author} {\bibfnamefont {W.}~\bibnamefont {Gehrke}},\ and\ \bibinfo {author} {\bibfnamefont {U.}~\bibnamefont {Helmbrecht}},\ }\bibfield  {title} {\enquote {\bibinfo {title} {{Non-equilibrium dynamics of a dissipative two-site Hubbard model simulated on IBM quantum computers}},}\ }\href {https://doi.org/10.1088/1751-8121/ac6bd0} {\bibfield  {journal} {\bibinfo  {journal} {J. Phys. A-Math.}\ }\textbf {\bibinfo {volume} {55}},\ \bibinfo {pages} {245302} (\bibinfo {year} {2022})},\ \Eprint {https://arxiv.org/abs/2011.11059} {2011.11059} \BibitemShut {NoStop}%
\bibitem [{\citenamefont {Schlimgen}\ \emph {et~al.}(2022{\natexlab{a}})\citenamefont {Schlimgen}, \citenamefont {Head-Marsden}, \citenamefont {Sager}, \citenamefont {Narang},\ and\ \citenamefont {Mazziotti}}]{Schlimgen.2022}%
  \BibitemOpen
  \bibfield  {author} {\bibinfo {author} {\bibfnamefont {A.~W.}\ \bibnamefont {Schlimgen}}, \bibinfo {author} {\bibfnamefont {K.}~\bibnamefont {Head-Marsden}}, \bibinfo {author} {\bibfnamefont {L.~M.}\ \bibnamefont {Sager}}, \bibinfo {author} {\bibfnamefont {P.}~\bibnamefont {Narang}},\ and\ \bibinfo {author} {\bibfnamefont {D.~A.}\ \bibnamefont {Mazziotti}},\ }\bibfield  {title} {\enquote {\bibinfo {title} {{Quantum simulation of the Lindblad equation using a unitary decomposition of operators}},}\ }\href {https://doi.org/10.1103/physrevresearch.4.023216} {\bibfield  {journal} {\bibinfo  {journal} {Phys. Rev. Research}\ }\textbf {\bibinfo {volume} {4}},\ \bibinfo {pages} {023216} (\bibinfo {year} {2022}{\natexlab{a}})}\BibitemShut {NoStop}%
\bibitem [{\citenamefont {Schlimgen}\ \emph {et~al.}(2022{\natexlab{b}})\citenamefont {Schlimgen}, \citenamefont {Head-Marsden}, \citenamefont {Sager-Smith}, \citenamefont {Narang},\ and\ \citenamefont {Mazziotti}}]{Schlimgen.2022lz7}%
  \BibitemOpen
  \bibfield  {author} {\bibinfo {author} {\bibfnamefont {A.~W.}\ \bibnamefont {Schlimgen}}, \bibinfo {author} {\bibfnamefont {K.}~\bibnamefont {Head-Marsden}}, \bibinfo {author} {\bibfnamefont {L.~M.}\ \bibnamefont {Sager-Smith}}, \bibinfo {author} {\bibfnamefont {P.}~\bibnamefont {Narang}},\ and\ \bibinfo {author} {\bibfnamefont {D.~A.}\ \bibnamefont {Mazziotti}},\ }\bibfield  {title} {\enquote {\bibinfo {title} {{Quantum state preparation and nonunitary evolution with diagonal operators}},}\ }\href {https://doi.org/10.1103/physreva.106.022414} {\bibfield  {journal} {\bibinfo  {journal} {Phys. Rev. A}\ }\textbf {\bibinfo {volume} {106}},\ \bibinfo {pages} {022414} (\bibinfo {year} {2022}{\natexlab{b}})},\ \Eprint {https://arxiv.org/abs/2205.02826} {2205.02826} \BibitemShut {NoStop}%
\bibitem [{\citenamefont {Liang}\ \emph {et~al.}(2023)\citenamefont {Liang}, \citenamefont {Lv}, \citenamefont {Wang},\ and\ \citenamefont {Fei}}]{Liang.2023}%
  \BibitemOpen
  \bibfield  {author} {\bibinfo {author} {\bibfnamefont {J.-M.}\ \bibnamefont {Liang}}, \bibinfo {author} {\bibfnamefont {Q.-Q.}\ \bibnamefont {Lv}}, \bibinfo {author} {\bibfnamefont {Z.-X.}\ \bibnamefont {Wang}},\ and\ \bibinfo {author} {\bibfnamefont {S.-M.}\ \bibnamefont {Fei}},\ }\bibfield  {title} {\enquote {\bibinfo {title} {{Assisted quantum simulation of open quantum systems}},}\ }\href {https://doi.org/10.1016/j.isci.2023.106306} {\bibfield  {journal} {\bibinfo  {journal} {iScience}\ }\textbf {\bibinfo {volume} {26}},\ \bibinfo {pages} {106306} (\bibinfo {year} {2023})},\ \Eprint {https://arxiv.org/abs/2302.13299} {2302.13299} \BibitemShut {NoStop}%
\bibitem [{\citenamefont {Lau}\ \emph {et~al.}(2023)\citenamefont {Lau}, \citenamefont {Lim}, \citenamefont {Bharti}, \citenamefont {Kwek},\ and\ \citenamefont {Vinjanampathy}}]{Lau.2023}%
  \BibitemOpen
  \bibfield  {author} {\bibinfo {author} {\bibfnamefont {J.~W.~Z.}\ \bibnamefont {Lau}}, \bibinfo {author} {\bibfnamefont {K.~H.}\ \bibnamefont {Lim}}, \bibinfo {author} {\bibfnamefont {K.}~\bibnamefont {Bharti}}, \bibinfo {author} {\bibfnamefont {L.-C.}\ \bibnamefont {Kwek}},\ and\ \bibinfo {author} {\bibfnamefont {S.}~\bibnamefont {Vinjanampathy}},\ }\bibfield  {title} {\enquote {\bibinfo {title} {{Convex Optimization for Nonequilibrium Steady States on a Hybrid Quantum Processor}},}\ }\href {https://doi.org/10.1103/physrevlett.130.240601} {\bibfield  {journal} {\bibinfo  {journal} {Phys. Rev. Lett.}\ }\textbf {\bibinfo {volume} {130}},\ \bibinfo {pages} {240601} (\bibinfo {year} {2023})},\ \Eprint {https://arxiv.org/abs/2204.03203} {2204.03203} \BibitemShut {NoStop}%
\bibitem [{\citenamefont {Wang}\ \emph {et~al.}(2023)\citenamefont {Wang}, \citenamefont {Mulvihill}, \citenamefont {Hu}, \citenamefont {Lyu}, \citenamefont {Shivpuje}, \citenamefont {Liu}, \citenamefont {Soley}, \citenamefont {Geva}, \citenamefont {Batista},\ and\ \citenamefont {Kais}}]{Wang.20239eb}%
  \BibitemOpen
  \bibfield  {author} {\bibinfo {author} {\bibfnamefont {Y.}~\bibnamefont {Wang}}, \bibinfo {author} {\bibfnamefont {E.}~\bibnamefont {Mulvihill}}, \bibinfo {author} {\bibfnamefont {Z.}~\bibnamefont {Hu}}, \bibinfo {author} {\bibfnamefont {N.}~\bibnamefont {Lyu}}, \bibinfo {author} {\bibfnamefont {S.}~\bibnamefont {Shivpuje}}, \bibinfo {author} {\bibfnamefont {Y.}~\bibnamefont {Liu}}, \bibinfo {author} {\bibfnamefont {M.~B.}\ \bibnamefont {Soley}}, \bibinfo {author} {\bibfnamefont {E.}~\bibnamefont {Geva}}, \bibinfo {author} {\bibfnamefont {V.~S.}\ \bibnamefont {Batista}},\ and\ \bibinfo {author} {\bibfnamefont {S.}~\bibnamefont {Kais}},\ }\bibfield  {title} {\enquote {\bibinfo {title} {{Simulating Open Quantum System Dynamics on NISQ Computers with Generalized Quantum Master Equations}},}\ }\href {https://doi.org/10.1021/acs.jctc.3c00316} {\bibfield  {journal} {\bibinfo  {journal} {J. Chem. Theory Comput.}\ }\textbf {\bibinfo {volume} {19}},\ \bibinfo {pages} {4851--4862} (\bibinfo {year} {2023})},\ \Eprint
  {https://arxiv.org/abs/2209.04956} {2209.04956} \BibitemShut {NoStop}%
\bibitem [{\citenamefont {Li}\ and\ \citenamefont {Wang}(2023)}]{Li.2023}%
  \BibitemOpen
  \bibfield  {author} {\bibinfo {author} {\bibfnamefont {X.}~\bibnamefont {Li}}\ and\ \bibinfo {author} {\bibfnamefont {C.}~\bibnamefont {Wang}},\ }\bibfield  {title} {\enquote {\bibinfo {title} {{Succinct Description and Efficient Simulation of Non-Markovian Open Quantum Systems}},}\ }\href {https://doi.org/10.1007/s00220-023-04638-4} {\bibfield  {journal} {\bibinfo  {journal} {Commun. Math. Phys.}\ }\textbf {\bibinfo {volume} {401}},\ \bibinfo {pages} {147--183} (\bibinfo {year} {2023})},\ \Eprint {https://arxiv.org/abs/2111.03240} {2111.03240} \BibitemShut {NoStop}%
\bibitem [{\citenamefont {Miessen}\ \emph {et~al.}(2023)\citenamefont {Miessen}, \citenamefont {Ollitrault}, \citenamefont {Tacchino},\ and\ \citenamefont {Tavernelli}}]{Miessen.2023}%
  \BibitemOpen
  \bibfield  {author} {\bibinfo {author} {\bibfnamefont {A.}~\bibnamefont {Miessen}}, \bibinfo {author} {\bibfnamefont {P.~J.}\ \bibnamefont {Ollitrault}}, \bibinfo {author} {\bibfnamefont {F.}~\bibnamefont {Tacchino}},\ and\ \bibinfo {author} {\bibfnamefont {I.}~\bibnamefont {Tavernelli}},\ }\bibfield  {title} {\enquote {\bibinfo {title} {{Quantum algorithms for quantum dynamics}},}\ }\href {https://doi.org/10.1038/s43588-022-00374-2} {\bibfield  {journal} {\bibinfo  {journal} {Nat. Comput. Sci.}\ }\textbf {\bibinfo {volume} {3}},\ \bibinfo {pages} {25--37} (\bibinfo {year} {2023})}\BibitemShut {NoStop}%
\bibitem [{\citenamefont {Peetz}\ \emph {et~al.}(2023)\citenamefont {Peetz}, \citenamefont {Smart}, \citenamefont {Tserkis},\ and\ \citenamefont {Narang}}]{Peetz.2023}%
  \BibitemOpen
  \bibfield  {author} {\bibinfo {author} {\bibfnamefont {J.}~\bibnamefont {Peetz}}, \bibinfo {author} {\bibfnamefont {S.~E.}\ \bibnamefont {Smart}}, \bibinfo {author} {\bibfnamefont {S.}~\bibnamefont {Tserkis}},\ and\ \bibinfo {author} {\bibfnamefont {P.}~\bibnamefont {Narang}},\ }\bibfield  {title} {\enquote {\bibinfo {title} {{Simulation of Open Quantum Systems via Low-Depth Convex Unitary Evolutions}},}\ }\href {https://doi.org/10.48550/arxiv.2307.14325} {\bibfield  {journal} {\bibinfo  {journal} {arXiv}\ } (\bibinfo {year} {2023}),\ 10.48550/arxiv.2307.14325},\ \Eprint {https://arxiv.org/abs/2307.14325} {2307.14325} \BibitemShut {NoStop}%
\bibitem [{\citenamefont {Régent}\ and\ \citenamefont {Rouchon}(2023)}]{Régent.2023}%
  \BibitemOpen
  \bibfield  {author} {\bibinfo {author} {\bibfnamefont {F.-M.~L.}\ \bibnamefont {Régent}}\ and\ \bibinfo {author} {\bibfnamefont {P.}~\bibnamefont {Rouchon}},\ }\bibfield  {title} {\enquote {\bibinfo {title} {{Adiabatic elimination for composite open quantum systems: reduced model formulation and numerical simulations}},}\ }\href {https://doi.org/10.48550/arxiv.2303.05089} {\bibfield  {journal} {\bibinfo  {journal} {arXiv}\ } (\bibinfo {year} {2023}),\ 10.48550/arxiv.2303.05089},\ \Eprint {https://arxiv.org/abs/2303.05089} {2303.05089} \BibitemShut {NoStop}%
\bibitem [{\citenamefont {Schlegel}, \citenamefont {Minganti},\ and\ \citenamefont {Savona}(2023)}]{Schlegel.2023}%
  \BibitemOpen
  \bibfield  {author} {\bibinfo {author} {\bibfnamefont {D.~S.}\ \bibnamefont {Schlegel}}, \bibinfo {author} {\bibfnamefont {F.}~\bibnamefont {Minganti}},\ and\ \bibinfo {author} {\bibfnamefont {V.}~\bibnamefont {Savona}},\ }\bibfield  {title} {\enquote {\bibinfo {title} {{Coherent-State Ladder Time-Dependent Variational Principle for Open Quantum Systems}},}\ }\href {https://doi.org/10.48550/arxiv.2306.13708} {\bibfield  {journal} {\bibinfo  {journal} {arXiv}\ } (\bibinfo {year} {2023}),\ 10.48550/arxiv.2306.13708},\ \Eprint {https://arxiv.org/abs/2306.13708} {2306.13708} \BibitemShut {NoStop}%
\bibitem [{\citenamefont {Zhou}, \citenamefont {Mao},\ and\ \citenamefont {Sun}(2023)}]{Zhou.2023nw3}%
  \BibitemOpen
  \bibfield  {author} {\bibinfo {author} {\bibfnamefont {H.}~\bibnamefont {Zhou}}, \bibinfo {author} {\bibfnamefont {R.}~\bibnamefont {Mao}},\ and\ \bibinfo {author} {\bibfnamefont {X.}~\bibnamefont {Sun}},\ }\bibfield  {title} {\enquote {\bibinfo {title} {{Hybrid algorithm simulating non-equilibrium steady states of an open quantum system}},}\ }\href {https://doi.org/10.48550/arxiv.2309.06665} {\bibfield  {journal} {\bibinfo  {journal} {arXiv}\ } (\bibinfo {year} {2023}),\ 10.48550/arxiv.2309.06665},\ \Eprint {https://arxiv.org/abs/2309.06665} {2309.06665} \BibitemShut {NoStop}%
\bibitem [{\citenamefont {Rossini}\ \emph {et~al.}(2023)\citenamefont {Rossini}, \citenamefont {Maile}, \citenamefont {Ankerhold},\ and\ \citenamefont {Donvil}}]{Rossini.2023}%
  \BibitemOpen
  \bibfield  {author} {\bibinfo {author} {\bibfnamefont {M.}~\bibnamefont {Rossini}}, \bibinfo {author} {\bibfnamefont {D.}~\bibnamefont {Maile}}, \bibinfo {author} {\bibfnamefont {J.}~\bibnamefont {Ankerhold}},\ and\ \bibinfo {author} {\bibfnamefont {B.~I.~C.}\ \bibnamefont {Donvil}},\ }\bibfield  {title} {\enquote {\bibinfo {title} {{Single-Qubit Error Mitigation by Simulating Non-Markovian Dynamics}},}\ }\href {https://doi.org/10.1103/physrevlett.131.110603} {\bibfield  {journal} {\bibinfo  {journal} {Phys. Rev. Lett.}\ }\textbf {\bibinfo {volume} {131}},\ \bibinfo {pages} {110603} (\bibinfo {year} {2023})},\ \Eprint {https://arxiv.org/abs/2303.03268} {2303.03268} \BibitemShut {NoStop}%
\bibitem [{\citenamefont {David}, \citenamefont {Sinayskiy},\ and\ \citenamefont {Petruccione}(2023)}]{David.2023}%
  \BibitemOpen
  \bibfield  {author} {\bibinfo {author} {\bibfnamefont {I.~J.}\ \bibnamefont {David}}, \bibinfo {author} {\bibfnamefont {I.}~\bibnamefont {Sinayskiy}},\ and\ \bibinfo {author} {\bibfnamefont {F.}~\bibnamefont {Petruccione}},\ }\bibfield  {title} {\enquote {\bibinfo {title} {{Digital Simulation of Single Qubit Markovian Open Quantum Systems: A Tutorial}},}\ }\href {https://doi.org/10.12743/quanta.v12i1.226} {\bibfield  {journal} {\bibinfo  {journal} {Quanta}\ }\textbf {\bibinfo {volume} {12}},\ \bibinfo {pages} {131--163} (\bibinfo {year} {2023})},\ \Eprint {https://arxiv.org/abs/2302.02953} {2302.02953} \BibitemShut {NoStop}%
\bibitem [{\citenamefont {Suri}\ \emph {et~al.}(2023{\natexlab{a}})\citenamefont {Suri}, \citenamefont {Barreto}, \citenamefont {Hadfield}, \citenamefont {Wiebe}, \citenamefont {Wudarski},\ and\ \citenamefont {Marshall}}]{Suri.2023}%
  \BibitemOpen
  \bibfield  {author} {\bibinfo {author} {\bibfnamefont {N.}~\bibnamefont {Suri}}, \bibinfo {author} {\bibfnamefont {J.}~\bibnamefont {Barreto}}, \bibinfo {author} {\bibfnamefont {S.}~\bibnamefont {Hadfield}}, \bibinfo {author} {\bibfnamefont {N.}~\bibnamefont {Wiebe}}, \bibinfo {author} {\bibfnamefont {F.}~\bibnamefont {Wudarski}},\ and\ \bibinfo {author} {\bibfnamefont {J.}~\bibnamefont {Marshall}},\ }\bibfield  {title} {\enquote {\bibinfo {title} {{Two-Unitary Decomposition Algorithm and Open Quantum System Simulation}},}\ }\href {https://doi.org/10.22331/q-2023-05-15-1002} {\bibfield  {journal} {\bibinfo  {journal} {Quantum}\ }\textbf {\bibinfo {volume} {7}},\ \bibinfo {pages} {1002} (\bibinfo {year} {2023}{\natexlab{a}})},\ \Eprint {https://arxiv.org/abs/2207.10007} {2207.10007} \BibitemShut {NoStop}%
\bibitem [{\citenamefont {Zhang}\ \emph {et~al.}(2023)\citenamefont {Zhang}, \citenamefont {Hu}, \citenamefont {Wang},\ and\ \citenamefont {Kais}}]{Zhang.2023z47}%
  \BibitemOpen
  \bibfield  {author} {\bibinfo {author} {\bibfnamefont {Y.}~\bibnamefont {Zhang}}, \bibinfo {author} {\bibfnamefont {Z.}~\bibnamefont {Hu}}, \bibinfo {author} {\bibfnamefont {Y.}~\bibnamefont {Wang}},\ and\ \bibinfo {author} {\bibfnamefont {S.}~\bibnamefont {Kais}},\ }\bibfield  {title} {\enquote {\bibinfo {title} {{Quantum Simulation of the Radical Pair Dynamics of the Avian Compass}},}\ }\href {https://doi.org/10.1021/acs.jpclett.2c03617} {\bibfield  {journal} {\bibinfo  {journal} {J. Phys. Chem. Lett.}\ }\textbf {\bibinfo {volume} {14}},\ \bibinfo {pages} {832--837} (\bibinfo {year} {2023})},\ \Eprint {https://arxiv.org/abs/2211.15427} {2211.15427} \BibitemShut {NoStop}%
\bibitem [{\citenamefont {Avdic}, \citenamefont {Sager-Smith},\ and\ \citenamefont {Mazziotti}(2023)}]{Avdic.2023}%
  \BibitemOpen
  \bibfield  {author} {\bibinfo {author} {\bibfnamefont {I.}~\bibnamefont {Avdic}}, \bibinfo {author} {\bibfnamefont {L.~M.}\ \bibnamefont {Sager-Smith}},\ and\ \bibinfo {author} {\bibfnamefont {D.~A.}\ \bibnamefont {Mazziotti}},\ }\bibfield  {title} {\enquote {\bibinfo {title} {{Open quantum system violates generalized Pauli constraints on quantum device}},}\ }\href {https://doi.org/10.1038/s42005-023-01295-w} {\bibfield  {journal} {\bibinfo  {journal} {Commun. phys.}\ }\textbf {\bibinfo {volume} {6}},\ \bibinfo {pages} {180} (\bibinfo {year} {2023})}\BibitemShut {NoStop}%
\bibitem [{\citenamefont {Guimarães}\ \emph {et~al.}(2023)\citenamefont {Guimarães}, \citenamefont {Lim}, \citenamefont {Vasilevskiy}, \citenamefont {Huelga},\ and\ \citenamefont {Plenio}}]{Guimarães.2023}%
  \BibitemOpen
  \bibfield  {author} {\bibinfo {author} {\bibfnamefont {J.~D.}\ \bibnamefont {Guimarães}}, \bibinfo {author} {\bibfnamefont {J.}~\bibnamefont {Lim}}, \bibinfo {author} {\bibfnamefont {M.~I.}\ \bibnamefont {Vasilevskiy}}, \bibinfo {author} {\bibfnamefont {S.~F.}\ \bibnamefont {Huelga}},\ and\ \bibinfo {author} {\bibfnamefont {M.~B.}\ \bibnamefont {Plenio}},\ }\bibfield  {title} {\enquote {\bibinfo {title} {{Noise-Assisted Digital Quantum Simulation of Open Systems Using Partial Probabilistic Error Cancellation}},}\ }\href {https://doi.org/10.1103/prxquantum.4.040329} {\bibfield  {journal} {\bibinfo  {journal} {PRX Quantum}\ }\textbf {\bibinfo {volume} {4}},\ \bibinfo {pages} {040329} (\bibinfo {year} {2023})},\ \Eprint {https://arxiv.org/abs/2302.14592} {2302.14592} \BibitemShut {NoStop}%
\bibitem [{\citenamefont {Santos}, \citenamefont {Song},\ and\ \citenamefont {Savona}(2024)}]{Santos.2024}%
  \BibitemOpen
  \bibfield  {author} {\bibinfo {author} {\bibfnamefont {S.}~\bibnamefont {Santos}}, \bibinfo {author} {\bibfnamefont {X.}~\bibnamefont {Song}},\ and\ \bibinfo {author} {\bibfnamefont {V.}~\bibnamefont {Savona}},\ }\bibfield  {title} {\enquote {\bibinfo {title} {{Low-Rank Variational Quantum Algorithm for the Dynamics of Open Quantum Systems}},}\ }\href {https://doi.org/10.48550/arxiv.2403.05908} {\bibfield  {journal} {\bibinfo  {journal} {arXiv}\ } (\bibinfo {year} {2024}),\ 10.48550/arxiv.2403.05908},\ \Eprint {https://arxiv.org/abs/2403.05908} {2403.05908} \BibitemShut {NoStop}%
\bibitem [{\citenamefont {Guimarães}, \citenamefont {Vasilevskiy},\ and\ \citenamefont {Barbosa}(2024)}]{Guimarães.2024}%
  \BibitemOpen
  \bibfield  {author} {\bibinfo {author} {\bibfnamefont {J.~D.}\ \bibnamefont {Guimarães}}, \bibinfo {author} {\bibfnamefont {M.~I.}\ \bibnamefont {Vasilevskiy}},\ and\ \bibinfo {author} {\bibfnamefont {L.~S.}\ \bibnamefont {Barbosa}},\ }\bibfield  {title} {\enquote {\bibinfo {title} {{Digital quantum simulation of non-perturbative dynamics of open systems with orthogonal polynomials}},}\ }\href {https://doi.org/10.22331/q-2024-02-05-1242} {\bibfield  {journal} {\bibinfo  {journal} {Quantum}\ }\textbf {\bibinfo {volume} {8}},\ \bibinfo {pages} {1242} (\bibinfo {year} {2024})},\ \Eprint {https://arxiv.org/abs/2203.14653} {2203.14653} \BibitemShut {NoStop}%
\bibitem [{\citenamefont {Luo}, \citenamefont {Lin},\ and\ \citenamefont {Gao}(2024)}]{Luo.2024}%
  \BibitemOpen
  \bibfield  {author} {\bibinfo {author} {\bibfnamefont {J.}~\bibnamefont {Luo}}, \bibinfo {author} {\bibfnamefont {K.}~\bibnamefont {Lin}},\ and\ \bibinfo {author} {\bibfnamefont {X.}~\bibnamefont {Gao}},\ }\bibfield  {title} {\enquote {\bibinfo {title} {{Variational Quantum Simulation of Lindblad Dynamics via Quantum State Diffusion}},}\ }\href {https://doi.org/10.1021/acs.jpclett.4c00576} {\bibfield  {journal} {\bibinfo  {journal} {J. Phys. Chem. Lett.}\ }\textbf {\bibinfo {volume} {15}},\ \bibinfo {pages} {3516--3522} (\bibinfo {year} {2024})}\BibitemShut {NoStop}%
\bibitem [{\citenamefont {Re}\ \emph {et~al.}(2024)\citenamefont {Re}, \citenamefont {Rost}, \citenamefont {Foss-Feig}, \citenamefont {Kemper},\ and\ \citenamefont {Freericks}}]{Re.2024}%
  \BibitemOpen
  \bibfield  {author} {\bibinfo {author} {\bibfnamefont {L.~D.}\ \bibnamefont {Re}}, \bibinfo {author} {\bibfnamefont {B.}~\bibnamefont {Rost}}, \bibinfo {author} {\bibfnamefont {M.}~\bibnamefont {Foss-Feig}}, \bibinfo {author} {\bibfnamefont {A.~F.}\ \bibnamefont {Kemper}},\ and\ \bibinfo {author} {\bibfnamefont {J.~K.}\ \bibnamefont {Freericks}},\ }\bibfield  {title} {\enquote {\bibinfo {title} {{Robust Measurements of n-Point Correlation Functions of Driven-Dissipative Quantum Systems on a Digital Quantum Computer}},}\ }\href {https://doi.org/10.1103/physrevlett.132.100601} {\bibfield  {journal} {\bibinfo  {journal} {Phys. Rev. Lett.}\ }\textbf {\bibinfo {volume} {132}},\ \bibinfo {pages} {100601} (\bibinfo {year} {2024})}\BibitemShut {NoStop}%
\bibitem [{\citenamefont {Kunold}(2024)}]{Kunold.2024}%
  \BibitemOpen
  \bibfield  {author} {\bibinfo {author} {\bibfnamefont {A.}~\bibnamefont {Kunold}},\ }\bibfield  {title} {\enquote {\bibinfo {title} {{Vectorization of the density matrix and quantum simulation of the von Neumann equation of time-dependent Hamiltonians}},}\ }\href {https://doi.org/10.1088/1402-4896/ad44f4} {\bibfield  {journal} {\bibinfo  {journal} {Phys. Scr.}\ }\textbf {\bibinfo {volume} {99}},\ \bibinfo {pages} {065111} (\bibinfo {year} {2024})}\BibitemShut {NoStop}%
\bibitem [{\citenamefont {Chen}\ \emph {et~al.}(2024)\citenamefont {Chen}, \citenamefont {Gomes}, \citenamefont {Niu},\ and\ \citenamefont {Jong}}]{Chen.2024}%
  \BibitemOpen
  \bibfield  {author} {\bibinfo {author} {\bibfnamefont {H.}~\bibnamefont {Chen}}, \bibinfo {author} {\bibfnamefont {N.}~\bibnamefont {Gomes}}, \bibinfo {author} {\bibfnamefont {S.}~\bibnamefont {Niu}},\ and\ \bibinfo {author} {\bibfnamefont {W.~A.~d.}\ \bibnamefont {Jong}},\ }\bibfield  {title} {\enquote {\bibinfo {title} {{Adaptive variational simulation for open quantum systems}},}\ }\href {https://doi.org/10.22331/q-2024-02-13-1252} {\bibfield  {journal} {\bibinfo  {journal} {Quantum}\ }\textbf {\bibinfo {volume} {8}},\ \bibinfo {pages} {1252} (\bibinfo {year} {2024})},\ \Eprint {https://arxiv.org/abs/2305.06915} {2305.06915} \BibitemShut {NoStop}%
\bibitem [{\citenamefont {Shivpuje}\ \emph {et~al.}(2024)\citenamefont {Shivpuje}, \citenamefont {Sajjan}, \citenamefont {Wang}, \citenamefont {Hu},\ and\ \citenamefont {Kais}}]{Shivpuje.2024}%
  \BibitemOpen
  \bibfield  {author} {\bibinfo {author} {\bibfnamefont {S.}~\bibnamefont {Shivpuje}}, \bibinfo {author} {\bibfnamefont {M.}~\bibnamefont {Sajjan}}, \bibinfo {author} {\bibfnamefont {Y.}~\bibnamefont {Wang}}, \bibinfo {author} {\bibfnamefont {Z.}~\bibnamefont {Hu}},\ and\ \bibinfo {author} {\bibfnamefont {S.}~\bibnamefont {Kais}},\ }\bibfield  {title} {\enquote {\bibinfo {title} {{Designing Variational Ansatz for Quantum‐Enabled Simulation of Non‐Unitary Dynamical Evolution ‐ An Excursion into Dicke Supperradiance}},}\ }\href {https://doi.org/10.1002/qute.202400088} {\bibfield  {journal} {\bibinfo  {journal} {Adv. Quantum Technol.}\ } (\bibinfo {year} {2024}),\ 10.1002/qute.202400088}\BibitemShut {NoStop}%
\bibitem [{\citenamefont {Basile}\ and\ \citenamefont {Pineda}(2024)}]{Basile.2024}%
  \BibitemOpen
  \bibfield  {author} {\bibinfo {author} {\bibfnamefont {T.}~\bibnamefont {Basile}}\ and\ \bibinfo {author} {\bibfnamefont {C.}~\bibnamefont {Pineda}},\ }\bibfield  {title} {\enquote {\bibinfo {title} {{Quantum simulation of Pauli channels and dynamical maps: Algorithm and implementation}},}\ }\href {https://doi.org/10.1371/journal.pone.0297210} {\bibfield  {journal} {\bibinfo  {journal} {PLOS ONE}\ }\textbf {\bibinfo {volume} {19}},\ \bibinfo {pages} {e0297210} (\bibinfo {year} {2024})},\ \Eprint {https://arxiv.org/abs/2308.00188} {2308.00188} \BibitemShut {NoStop}%
\bibitem [{\citenamefont {Seneviratne}, \citenamefont {Walters},\ and\ \citenamefont {Wang}(2024)}]{Seneviratne.2024}%
  \BibitemOpen
  \bibfield  {author} {\bibinfo {author} {\bibfnamefont {A.}~\bibnamefont {Seneviratne}}, \bibinfo {author} {\bibfnamefont {P.~L.}\ \bibnamefont {Walters}},\ and\ \bibinfo {author} {\bibfnamefont {F.}~\bibnamefont {Wang}},\ }\bibfield  {title} {\enquote {\bibinfo {title} {{Exact Non-Markovian Quantum Dynamics on the NISQ Device Using Kraus Operators}},}\ }\href {https://doi.org/10.1021/acsomega.3c09720} {\bibfield  {journal} {\bibinfo  {journal} {ACS Omega}\ }\textbf {\bibinfo {volume} {9}},\ \bibinfo {pages} {9666--9675} (\bibinfo {year} {2024})}\BibitemShut {NoStop}%
\bibitem [{\citenamefont {Ding}, \citenamefont {Li},\ and\ \citenamefont {Lin}(2024)}]{Ding.2024}%
  \BibitemOpen
  \bibfield  {author} {\bibinfo {author} {\bibfnamefont {Z.}~\bibnamefont {Ding}}, \bibinfo {author} {\bibfnamefont {X.}~\bibnamefont {Li}},\ and\ \bibinfo {author} {\bibfnamefont {L.}~\bibnamefont {Lin}},\ }\bibfield  {title} {\enquote {\bibinfo {title} {{Simulating Open Quantum Systems Using Hamiltonian Simulations}},}\ }\href {https://doi.org/10.1103/prxquantum.5.020332} {\bibfield  {journal} {\bibinfo  {journal} {PRX Quantum}\ }\textbf {\bibinfo {volume} {5}},\ \bibinfo {pages} {020332} (\bibinfo {year} {2024})}\BibitemShut {NoStop}%
\bibitem [{\citenamefont {Watad}\ and\ \citenamefont {Lindner}(2024)}]{Watad.2024}%
  \BibitemOpen
  \bibfield  {author} {\bibinfo {author} {\bibfnamefont {T.~M.}\ \bibnamefont {Watad}}\ and\ \bibinfo {author} {\bibfnamefont {N.~H.}\ \bibnamefont {Lindner}},\ }\bibfield  {title} {\enquote {\bibinfo {title} {{Variational quantum algorithms for simulation of Lindblad dynamics}},}\ }\href {https://doi.org/10.1088/2058-9565/ad17d8} {\bibfield  {journal} {\bibinfo  {journal} {Quantum Sci. Technol.}\ }\textbf {\bibinfo {volume} {9}},\ \bibinfo {pages} {025015} (\bibinfo {year} {2024})}\BibitemShut {NoStop}%
\bibitem [{\citenamefont {Schlimgen}\ \emph {et~al.}(2022{\natexlab{c}})\citenamefont {Schlimgen}, \citenamefont {Head-Marsden}, \citenamefont {Sager-Smith}, \citenamefont {Narang},\ and\ \citenamefont {Mazziotti}}]{schlimgen2022}%
  \BibitemOpen
  \bibfield  {author} {\bibinfo {author} {\bibfnamefont {A.~W.}\ \bibnamefont {Schlimgen}}, \bibinfo {author} {\bibfnamefont {K.}~\bibnamefont {Head-Marsden}}, \bibinfo {author} {\bibfnamefont {L.~M.}\ \bibnamefont {Sager-Smith}}, \bibinfo {author} {\bibfnamefont {P.}~\bibnamefont {Narang}},\ and\ \bibinfo {author} {\bibfnamefont {D.~A.}\ \bibnamefont {Mazziotti}},\ }\href@noop {} {\enquote {\bibinfo {title} {Quantum simulation of open quantum systems using density-matrix purification},}\ } (\bibinfo {year} {2022}{\natexlab{c}}),\ \Eprint {https://arxiv.org/abs/2207.07112} {arXiv:2207.07112 [quant-ph]} \BibitemShut {NoStop}%
\bibitem [{\citenamefont {Shushkov}\ and\ \citenamefont {Miller}(2019)}]{10.1063/1.5121749}%
  \BibitemOpen
  \bibfield  {author} {\bibinfo {author} {\bibfnamefont {P.}~\bibnamefont {Shushkov}}\ and\ \bibinfo {author} {\bibfnamefont {I.}~\bibnamefont {Miller}, \bibfnamefont {Thomas~F.}},\ }\bibfield  {title} {\enquote {\bibinfo {title} {{Real-time density-matrix coupled-cluster approach for closed and open systems at finite temperature}},}\ }\href {https://doi.org/10.1063/1.5121749} {\bibfield  {journal} {\bibinfo  {journal} {J. Chem. Phys.}\ }\textbf {\bibinfo {volume} {151}},\ \bibinfo {pages} {134107} (\bibinfo {year} {2019})},\ \Eprint {https://arxiv.org/abs/https://pubs.aip.org/aip/jcp/article-pdf/doi/10.1063/1.5121749/16717590/134107\_1\_online.pdf} {https://pubs.aip.org/aip/jcp/article-pdf/doi/10.1063/1.5121749/16717590/134107\_1\_online.pdf} \BibitemShut {NoStop}%
\bibitem [{\citenamefont {Benavides-Riveros}\ \emph {et~al.}(2022)\citenamefont {Benavides-Riveros}, \citenamefont {Chen}, \citenamefont {Schilling}, \citenamefont {Mantilla},\ and\ \citenamefont {Pittalis}}]{PhysRevLett.129.066401}%
  \BibitemOpen
  \bibfield  {author} {\bibinfo {author} {\bibfnamefont {C.~L.}\ \bibnamefont {Benavides-Riveros}}, \bibinfo {author} {\bibfnamefont {L.}~\bibnamefont {Chen}}, \bibinfo {author} {\bibfnamefont {C.}~\bibnamefont {Schilling}}, \bibinfo {author} {\bibfnamefont {S.}~\bibnamefont {Mantilla}},\ and\ \bibinfo {author} {\bibfnamefont {S.}~\bibnamefont {Pittalis}},\ }\bibfield  {title} {\enquote {\bibinfo {title} {Excitations of quantum many-body systems via purified ensembles: A unitary-coupled-cluster-based approach},}\ }\href {https://doi.org/10.1103/PhysRevLett.129.066401} {\bibfield  {journal} {\bibinfo  {journal} {Phys. Rev. Lett.}\ }\textbf {\bibinfo {volume} {129}},\ \bibinfo {pages} {066401} (\bibinfo {year} {2022})}\BibitemShut {NoStop}%
\bibitem [{\citenamefont {TAKAHASHI}\ and\ \citenamefont {UMEZAWA}(1996)}]{doi:10.1142/S0217979296000817}%
  \BibitemOpen
  \bibfield  {author} {\bibinfo {author} {\bibfnamefont {Y.}~\bibnamefont {TAKAHASHI}}\ and\ \bibinfo {author} {\bibfnamefont {H.}~\bibnamefont {UMEZAWA}},\ }\bibfield  {title} {\enquote {\bibinfo {title} {Thermo field dynamics},}\ }\href {https://doi.org/10.1142/S0217979296000817} {\bibfield  {journal} {\bibinfo  {journal} {Int. J. Mod. Phys. B}\ }\textbf {\bibinfo {volume} {10}},\ \bibinfo {pages} {1755--1805} (\bibinfo {year} {1996})},\ \Eprint {https://arxiv.org/abs/https://doi.org/10.1142/S0217979296000817} {https://doi.org/10.1142/S0217979296000817} \BibitemShut {NoStop}%
\bibitem [{\citenamefont {Wilde}(2013)}]{wilde2013}%
  \BibitemOpen
  \bibfield  {author} {\bibinfo {author} {\bibfnamefont {M.~M.}\ \bibnamefont {Wilde}},\ }\href@noop {} {\emph {\bibinfo {title} {Quantum Information Theory}}}\ (\bibinfo  {publisher} {Cambridge university press},\ \bibinfo {year} {2013})\BibitemShut {NoStop}%
\bibitem [{\citenamefont {Nielsen}\ and\ \citenamefont {Chuang}(2010)}]{nielsen2010}%
  \BibitemOpen
  \bibfield  {author} {\bibinfo {author} {\bibfnamefont {M.~A.}\ \bibnamefont {Nielsen}}\ and\ \bibinfo {author} {\bibfnamefont {I.~L.}\ \bibnamefont {Chuang}},\ }\href@noop {} {\emph {\bibinfo {title} {Quantum computation and quantum information}}}\ (\bibinfo  {publisher} {Cambridge university press},\ \bibinfo {year} {2010})\BibitemShut {NoStop}%
\bibitem [{\citenamefont {Viennot}(2018)}]{viennot2018}%
  \BibitemOpen
  \bibfield  {author} {\bibinfo {author} {\bibfnamefont {D.}~\bibnamefont {Viennot}},\ }\bibfield  {title} {\enquote {\bibinfo {title} {Purification of {Lindblad} dynamics, geometry of mixed states and geometric phases},}\ }\href {https://doi.org/https://doi.org/10.1016/j.geomphys.2018.06.019} {\bibfield  {journal} {\bibinfo  {journal} {J. Geom. Phys.}\ }\textbf {\bibinfo {volume} {133}},\ \bibinfo {pages} {42--70} (\bibinfo {year} {2018})}\BibitemShut {NoStop}%
\bibitem [{\citenamefont {Kleinmann}\ \emph {et~al.}(2006)\citenamefont {Kleinmann}, \citenamefont {Kampermann}, \citenamefont {Meyer},\ and\ \citenamefont {Bru\ss{}}}]{PhysRevA.73.062309}%
  \BibitemOpen
  \bibfield  {author} {\bibinfo {author} {\bibfnamefont {M.}~\bibnamefont {Kleinmann}}, \bibinfo {author} {\bibfnamefont {H.}~\bibnamefont {Kampermann}}, \bibinfo {author} {\bibfnamefont {T.}~\bibnamefont {Meyer}},\ and\ \bibinfo {author} {\bibfnamefont {D.}~\bibnamefont {Bru\ss{}}},\ }\bibfield  {title} {\enquote {\bibinfo {title} {Physical purification of quantum states},}\ }\href {https://doi.org/10.1103/PhysRevA.73.062309} {\bibfield  {journal} {\bibinfo  {journal} {Phys. Rev. A}\ }\textbf {\bibinfo {volume} {73}},\ \bibinfo {pages} {062309} (\bibinfo {year} {2006})}\BibitemShut {NoStop}%
\bibitem [{\citenamefont {Rivas}\ and\ \citenamefont {Huelga}(2012)}]{rivas2012open}%
  \BibitemOpen
  \bibfield  {author} {\bibinfo {author} {\bibfnamefont {A.}~\bibnamefont {Rivas}}\ and\ \bibinfo {author} {\bibfnamefont {S.~F.}\ \bibnamefont {Huelga}},\ }\href@noop {} {\emph {\bibinfo {title} {Open quantum systems}}},\ Vol.~\bibinfo {volume} {10}\ (\bibinfo  {publisher} {Springer},\ \bibinfo {year} {2012})\BibitemShut {NoStop}%
\bibitem [{\citenamefont {Jordan}, \citenamefont {Shaji},\ and\ \citenamefont {Sudarshan}(2004)}]{jordan2004}%
  \BibitemOpen
  \bibfield  {author} {\bibinfo {author} {\bibfnamefont {T.~F.}\ \bibnamefont {Jordan}}, \bibinfo {author} {\bibfnamefont {A.}~\bibnamefont {Shaji}},\ and\ \bibinfo {author} {\bibfnamefont {E.~C.~G.}\ \bibnamefont {Sudarshan}},\ }\bibfield  {title} {\enquote {\bibinfo {title} {Dynamics of initially entangled open quantum systems},}\ }\href {https://doi.org/10.1103/PhysRevA.70.052110} {\bibfield  {journal} {\bibinfo  {journal} {Phys. Rev. A}\ }\textbf {\bibinfo {volume} {70}},\ \bibinfo {pages} {052110} (\bibinfo {year} {2004})}\BibitemShut {NoStop}%
\bibitem [{\citenamefont {Ángel Rivas}, \citenamefont {Huelga},\ and\ \citenamefont {Plenio}(2014)}]{Rivas2014}%
  \BibitemOpen
  \bibfield  {author} {\bibinfo {author} {\bibnamefont {Ángel Rivas}}, \bibinfo {author} {\bibfnamefont {S.~F.}\ \bibnamefont {Huelga}},\ and\ \bibinfo {author} {\bibfnamefont {M.~B.}\ \bibnamefont {Plenio}},\ }\bibfield  {title} {\enquote {\bibinfo {title} {Quantum non-markovianity: characterization, quantification and detection},}\ }\href {https://doi.org/10.1088/0034-4885/77/9/094001} {\bibfield  {journal} {\bibinfo  {journal} {Rep. Prog. Phys.}\ }\textbf {\bibinfo {volume} {77}},\ \bibinfo {pages} {094001} (\bibinfo {year} {2014})}\BibitemShut {NoStop}%
\bibitem [{\citenamefont {Pechukas}(1994)}]{pechukas1994}%
  \BibitemOpen
  \bibfield  {author} {\bibinfo {author} {\bibfnamefont {P.}~\bibnamefont {Pechukas}},\ }\bibfield  {title} {\enquote {\bibinfo {title} {Reduced dynamics need not be completely positive},}\ }\href {https://doi.org/10.1103/PhysRevLett.73.1060} {\bibfield  {journal} {\bibinfo  {journal} {Phys. Rev. Lett.}\ }\textbf {\bibinfo {volume} {73}},\ \bibinfo {pages} {1060--1062} (\bibinfo {year} {1994})}\BibitemShut {NoStop}%
\bibitem [{\citenamefont {Carteret}, \citenamefont {Terno},\ and\ \citenamefont {\ifmmode~\dot{Z}\else \.{Z}\fi{}yczkowski}(2008)}]{carteret2008}%
  \BibitemOpen
  \bibfield  {author} {\bibinfo {author} {\bibfnamefont {H.~A.}\ \bibnamefont {Carteret}}, \bibinfo {author} {\bibfnamefont {D.~R.}\ \bibnamefont {Terno}},\ and\ \bibinfo {author} {\bibfnamefont {K.}~\bibnamefont {\ifmmode~\dot{Z}\else \.{Z}\fi{}yczkowski}},\ }\bibfield  {title} {\enquote {\bibinfo {title} {Dynamics beyond completely positive maps: Some properties and applications},}\ }\href {https://doi.org/10.1103/PhysRevA.77.042113} {\bibfield  {journal} {\bibinfo  {journal} {Phys. Rev. A}\ }\textbf {\bibinfo {volume} {77}},\ \bibinfo {pages} {042113} (\bibinfo {year} {2008})}\BibitemShut {NoStop}%
\bibitem [{\citenamefont {Alicki}(1995)}]{alicki1995}%
  \BibitemOpen
  \bibfield  {author} {\bibinfo {author} {\bibfnamefont {R.}~\bibnamefont {Alicki}},\ }\bibfield  {title} {\enquote {\bibinfo {title} {Comment on ``reduced dynamics need not be completely positive''},}\ }\href {https://doi.org/10.1103/PhysRevLett.75.3020} {\bibfield  {journal} {\bibinfo  {journal} {Phys. Rev. Lett.}\ }\textbf {\bibinfo {volume} {75}},\ \bibinfo {pages} {3020--3020} (\bibinfo {year} {1995})}\BibitemShut {NoStop}%
\bibitem [{\citenamefont {Pechukas}(1995)}]{pechukas1995}%
  \BibitemOpen
  \bibfield  {author} {\bibinfo {author} {\bibfnamefont {P.}~\bibnamefont {Pechukas}},\ }\bibfield  {title} {\enquote {\bibinfo {title} {Pechukas replies:},}\ }\href {https://doi.org/10.1103/PhysRevLett.75.3021} {\bibfield  {journal} {\bibinfo  {journal} {Phys. Rev. Lett.}\ }\textbf {\bibinfo {volume} {75}},\ \bibinfo {pages} {3021--3021} (\bibinfo {year} {1995})}\BibitemShut {NoStop}%
\bibitem [{\citenamefont {Sargolzahi}(2020)}]{sargolzahi2020}%
  \BibitemOpen
  \bibfield  {author} {\bibinfo {author} {\bibfnamefont {I.}~\bibnamefont {Sargolzahi}},\ }\bibfield  {title} {\enquote {\bibinfo {title} {Positivity of the assignment map implies complete positivity of the reduced dynamics},}\ }\href {https://doi.org/10.1007/s11128-020-02810-6} {\bibfield  {journal} {\bibinfo  {journal} {Quantum Inf. Process.}\ }\textbf {\bibinfo {volume} {19}},\ \bibinfo {pages} {310} (\bibinfo {year} {2020})}\BibitemShut {NoStop}%
\bibitem [{\citenamefont {Suri}\ \emph {et~al.}(2023{\natexlab{b}})\citenamefont {Suri}, \citenamefont {Barreto}, \citenamefont {Hadfield}, \citenamefont {Wiebe}, \citenamefont {Wudarski},\ and\ \citenamefont {Marshall}}]{Suri2023}%
  \BibitemOpen
  \bibfield  {author} {\bibinfo {author} {\bibfnamefont {N.}~\bibnamefont {Suri}}, \bibinfo {author} {\bibfnamefont {J.}~\bibnamefont {Barreto}}, \bibinfo {author} {\bibfnamefont {S.}~\bibnamefont {Hadfield}}, \bibinfo {author} {\bibfnamefont {N.}~\bibnamefont {Wiebe}}, \bibinfo {author} {\bibfnamefont {F.}~\bibnamefont {Wudarski}},\ and\ \bibinfo {author} {\bibfnamefont {J.}~\bibnamefont {Marshall}},\ }\bibfield  {title} {\enquote {\bibinfo {title} {Two-{U}nitary {D}ecomposition {A}lgorithm and {O}pen {Q}uantum {S}ystem {S}imulation},}\ }\href {https://doi.org/10.22331/q-2023-05-15-1002} {\bibfield  {journal} {\bibinfo  {journal} {{Quantum}}\ }\textbf {\bibinfo {volume} {7}},\ \bibinfo {pages} {1002} (\bibinfo {year} {2023}{\natexlab{b}})}\BibitemShut {NoStop}%
\bibitem [{\citenamefont {Ticozzi}\ and\ \citenamefont {Viola}(2017)}]{ticozzi2017quantum}%
  \BibitemOpen
  \bibfield  {author} {\bibinfo {author} {\bibfnamefont {F.}~\bibnamefont {Ticozzi}}\ and\ \bibinfo {author} {\bibfnamefont {L.}~\bibnamefont {Viola}},\ }\bibfield  {title} {\enquote {\bibinfo {title} {Quantum and classical resources for unitary design of open-system evolutions},}\ }\href@noop {} {\bibfield  {journal} {\bibinfo  {journal} {Quantum Sci. Technol.}\ }\textbf {\bibinfo {volume} {2}},\ \bibinfo {pages} {034001} (\bibinfo {year} {2017})}\BibitemShut {NoStop}%
\bibitem [{\citenamefont {Ando}\ and\ \citenamefont {Choi}(1986)}]{ando1986}%
  \BibitemOpen
  \bibfield  {author} {\bibinfo {author} {\bibfnamefont {T.}~\bibnamefont {Ando}}\ and\ \bibinfo {author} {\bibfnamefont {M.-D.}\ \bibnamefont {Choi}},\ }\bibfield  {title} {\enquote {\bibinfo {title} {Non-{Linear} {Completely} {Positive} {Maps}},}\ }in\ \href {https://doi.org/https://doi.org/10.1016/S0304-0208(08)71944-3} {\emph {\bibinfo {booktitle} {Aspects of {Positivity} in {Functional} {Analysis}}}},\ \bibinfo {series} {North-{Holland} {Mathematics} {Studies}}, Vol.\ \bibinfo {volume} {122},\ \bibinfo {editor} {edited by\ \bibinfo {editor} {\bibfnamefont {R.}~\bibnamefont {Nagel}}, \bibinfo {editor} {\bibfnamefont {U.}~\bibnamefont {Schloterbeck}},\ and\ \bibinfo {editor} {\bibfnamefont {M.~P.~H.}\ \bibnamefont {Wolff}}}\ (\bibinfo  {publisher} {North-Holland},\ \bibinfo {year} {1986})\ pp.\ \bibinfo {pages} {3--13},\ \bibinfo {note} {iSSN: 0304-0208}\BibitemShut {NoStop}%
\bibitem [{\citenamefont {Cook}, \citenamefont {Ko},\ and\ \citenamefont {Whaley}(2022)}]{cook2022quantum}%
  \BibitemOpen
  \bibfield  {author} {\bibinfo {author} {\bibfnamefont {R.~L.}\ \bibnamefont {Cook}}, \bibinfo {author} {\bibfnamefont {L.}~\bibnamefont {Ko}},\ and\ \bibinfo {author} {\bibfnamefont {K.~B.}\ \bibnamefont {Whaley}},\ }\href@noop {} {\enquote {\bibinfo {title} {A quantum trajectory picture of single photon absorption and energy transport in photosystem ii},}\ } (\bibinfo {year} {2022}),\ \Eprint {https://arxiv.org/abs/2110.13811} {arXiv:2110.13811 [physics.bio-ph]} \BibitemShut {NoStop}%
\bibitem [{\citenamefont {Engel}\ \emph {et~al.}(2007)\citenamefont {Engel}, \citenamefont {Calhoun}, \citenamefont {Read}, \citenamefont {Ahn}, \citenamefont {Man{\v{c}}al}, \citenamefont {Cheng}, \citenamefont {Blankenship},\ and\ \citenamefont {Fleming}}]{engel2007evidence}%
  \BibitemOpen
  \bibfield  {author} {\bibinfo {author} {\bibfnamefont {G.~S.}\ \bibnamefont {Engel}}, \bibinfo {author} {\bibfnamefont {T.~R.}\ \bibnamefont {Calhoun}}, \bibinfo {author} {\bibfnamefont {E.~L.}\ \bibnamefont {Read}}, \bibinfo {author} {\bibfnamefont {T.-K.}\ \bibnamefont {Ahn}}, \bibinfo {author} {\bibfnamefont {T.}~\bibnamefont {Man{\v{c}}al}}, \bibinfo {author} {\bibfnamefont {Y.-C.}\ \bibnamefont {Cheng}}, \bibinfo {author} {\bibfnamefont {R.~E.}\ \bibnamefont {Blankenship}},\ and\ \bibinfo {author} {\bibfnamefont {G.~R.}\ \bibnamefont {Fleming}},\ }\bibfield  {title} {\enquote {\bibinfo {title} {Evidence for wavelike energy transfer through quantum coherence in photosynthetic systems},}\ }\href@noop {} {\bibfield  {journal} {\bibinfo  {journal} {Nature}\ }\textbf {\bibinfo {volume} {446}},\ \bibinfo {pages} {782--786} (\bibinfo {year} {2007})}\BibitemShut {NoStop}%
\bibitem [{\citenamefont {Schouten}, \citenamefont {Sager-Smith},\ and\ \citenamefont {Mazziotti}(2023)}]{Schouten.2023cha}%
  \BibitemOpen
  \bibfield  {author} {\bibinfo {author} {\bibfnamefont {A.~O.}\ \bibnamefont {Schouten}}, \bibinfo {author} {\bibfnamefont {L.~M.}\ \bibnamefont {Sager-Smith}},\ and\ \bibinfo {author} {\bibfnamefont {D.~A.}\ \bibnamefont {Mazziotti}},\ }\bibfield  {title} {\enquote {\bibinfo {title} {{Exciton-Condensate-Like Amplification of Energy Transport in Light Harvesting}},}\ }\href {https://doi.org/10.1103/prxenergy.2.023002} {\bibfield  {journal} {\bibinfo  {journal} {PRX Energy}\ }\textbf {\bibinfo {volume} {2}},\ \bibinfo {pages} {023002} (\bibinfo {year} {2023})}\BibitemShut {NoStop}%
\bibitem [{\citenamefont {Brumer}\ and\ \citenamefont {Shapiro}(1989)}]{Brumer1989}%
  \BibitemOpen
  \bibfield  {author} {\bibinfo {author} {\bibfnamefont {P.}~\bibnamefont {Brumer}}\ and\ \bibinfo {author} {\bibfnamefont {M.}~\bibnamefont {Shapiro}},\ }\bibfield  {title} {\enquote {\bibinfo {title} {Coherence chemistry: controlling chemical reactions [with lasers]},}\ }\href {https://doi.org/10.1021/ar00168a001} {\bibfield  {journal} {\bibinfo  {journal} {Acc. Chem. Res.}\ }\textbf {\bibinfo {volume} {22}},\ \bibinfo {pages} {407--413} (\bibinfo {year} {1989})},\ \Eprint {https://arxiv.org/abs/https://doi.org/10.1021/ar00168a001} {https://doi.org/10.1021/ar00168a001} \BibitemShut {NoStop}%
\bibitem [{\citenamefont {Javadi-Abhari}\ \emph {et~al.}(2024)\citenamefont {Javadi-Abhari}, \citenamefont {Treinish}, \citenamefont {Krsulich}, \citenamefont {Wood}, \citenamefont {Lishman}, \citenamefont {Gacon}, \citenamefont {Martiel}, \citenamefont {Nation}, \citenamefont {Bishop}, \citenamefont {Cross}, \citenamefont {Johnson},\ and\ \citenamefont {Gambetta}}]{qiskit2024}%
  \BibitemOpen
  \bibfield  {author} {\bibinfo {author} {\bibfnamefont {A.}~\bibnamefont {Javadi-Abhari}}, \bibinfo {author} {\bibfnamefont {M.}~\bibnamefont {Treinish}}, \bibinfo {author} {\bibfnamefont {K.}~\bibnamefont {Krsulich}}, \bibinfo {author} {\bibfnamefont {C.~J.}\ \bibnamefont {Wood}}, \bibinfo {author} {\bibfnamefont {J.}~\bibnamefont {Lishman}}, \bibinfo {author} {\bibfnamefont {J.}~\bibnamefont {Gacon}}, \bibinfo {author} {\bibfnamefont {S.}~\bibnamefont {Martiel}}, \bibinfo {author} {\bibfnamefont {P.~D.}\ \bibnamefont {Nation}}, \bibinfo {author} {\bibfnamefont {L.~S.}\ \bibnamefont {Bishop}}, \bibinfo {author} {\bibfnamefont {A.~W.}\ \bibnamefont {Cross}}, \bibinfo {author} {\bibfnamefont {B.~R.}\ \bibnamefont {Johnson}},\ and\ \bibinfo {author} {\bibfnamefont {J.~M.}\ \bibnamefont {Gambetta}},\ }\href {https://doi.org/10.48550/arXiv.2405.08810} {\enquote {\bibinfo {title} {Quantum computing with {Q}iskit},}\ } (\bibinfo {year} {2024}),\ \Eprint {https://arxiv.org/abs/2405.08810} {arXiv:2405.08810
  [quant-ph]} \BibitemShut {NoStop}%
\bibitem [{\citenamefont {Plesch}\ and\ \citenamefont {Brukner}(2011)}]{plesch2011quantum}%
  \BibitemOpen
  \bibfield  {author} {\bibinfo {author} {\bibfnamefont {M.}~\bibnamefont {Plesch}}\ and\ \bibinfo {author} {\bibfnamefont {{\v{C}}.}~\bibnamefont {Brukner}},\ }\bibfield  {title} {\enquote {\bibinfo {title} {Quantum-state preparation with universal gate decompositions},}\ }\href@noop {} {\bibfield  {journal} {\bibinfo  {journal} {Phys. Rev. A}\ }\textbf {\bibinfo {volume} {83}},\ \bibinfo {pages} {032302} (\bibinfo {year} {2011})}\BibitemShut {NoStop}%
\bibitem [{\citenamefont {Virtanen}\ \emph {et~al.}(2020)\citenamefont {Virtanen}, \citenamefont {Gommers}, \citenamefont {Oliphant}, \citenamefont {Haberland}, \citenamefont {Reddy}, \citenamefont {Cournapeau}, \citenamefont {Burovski}, \citenamefont {Peterson}, \citenamefont {Weckesser}, \citenamefont {Bright}, \citenamefont {{van der Walt}}, \citenamefont {Brett}, \citenamefont {Wilson}, \citenamefont {Millman}, \citenamefont {Mayorov}, \citenamefont {Nelson}, \citenamefont {Jones}, \citenamefont {Kern}, \citenamefont {Larson}, \citenamefont {Carey}, \citenamefont {Polat}, \citenamefont {Feng}, \citenamefont {Moore}, \citenamefont {{VanderPlas}}, \citenamefont {Laxalde}, \citenamefont {Perktold}, \citenamefont {Cimrman}, \citenamefont {Henriksen}, \citenamefont {Quintero}, \citenamefont {Harris}, \citenamefont {Archibald}, \citenamefont {Ribeiro}, \citenamefont {Pedregosa}, \citenamefont {{van Mulbregt}},\ and\ \citenamefont {{SciPy 1.0 Contributors}}}]{2020SciPy-NMeth}%
  \BibitemOpen
  \bibfield  {author} {\bibinfo {author} {\bibfnamefont {P.}~\bibnamefont {Virtanen}}, \bibinfo {author} {\bibfnamefont {R.}~\bibnamefont {Gommers}}, \bibinfo {author} {\bibfnamefont {T.~E.}\ \bibnamefont {Oliphant}}, \bibinfo {author} {\bibfnamefont {M.}~\bibnamefont {Haberland}}, \bibinfo {author} {\bibfnamefont {T.}~\bibnamefont {Reddy}}, \bibinfo {author} {\bibfnamefont {D.}~\bibnamefont {Cournapeau}}, \bibinfo {author} {\bibfnamefont {E.}~\bibnamefont {Burovski}}, \bibinfo {author} {\bibfnamefont {P.}~\bibnamefont {Peterson}}, \bibinfo {author} {\bibfnamefont {W.}~\bibnamefont {Weckesser}}, \bibinfo {author} {\bibfnamefont {J.}~\bibnamefont {Bright}}, \bibinfo {author} {\bibfnamefont {S.~J.}\ \bibnamefont {{van der Walt}}}, \bibinfo {author} {\bibfnamefont {M.}~\bibnamefont {Brett}}, \bibinfo {author} {\bibfnamefont {J.}~\bibnamefont {Wilson}}, \bibinfo {author} {\bibfnamefont {K.~J.}\ \bibnamefont {Millman}}, \bibinfo {author} {\bibfnamefont {N.}~\bibnamefont {Mayorov}}, \bibinfo {author} {\bibfnamefont
  {A.~R.~J.}\ \bibnamefont {Nelson}}, \bibinfo {author} {\bibfnamefont {E.}~\bibnamefont {Jones}}, \bibinfo {author} {\bibfnamefont {R.}~\bibnamefont {Kern}}, \bibinfo {author} {\bibfnamefont {E.}~\bibnamefont {Larson}}, \bibinfo {author} {\bibfnamefont {C.~J.}\ \bibnamefont {Carey}}, \bibinfo {author} {\bibfnamefont {{\.I}.}~\bibnamefont {Polat}}, \bibinfo {author} {\bibfnamefont {Y.}~\bibnamefont {Feng}}, \bibinfo {author} {\bibfnamefont {E.~W.}\ \bibnamefont {Moore}}, \bibinfo {author} {\bibfnamefont {J.}~\bibnamefont {{VanderPlas}}}, \bibinfo {author} {\bibfnamefont {D.}~\bibnamefont {Laxalde}}, \bibinfo {author} {\bibfnamefont {J.}~\bibnamefont {Perktold}}, \bibinfo {author} {\bibfnamefont {R.}~\bibnamefont {Cimrman}}, \bibinfo {author} {\bibfnamefont {I.}~\bibnamefont {Henriksen}}, \bibinfo {author} {\bibfnamefont {E.~A.}\ \bibnamefont {Quintero}}, \bibinfo {author} {\bibfnamefont {C.~R.}\ \bibnamefont {Harris}}, \bibinfo {author} {\bibfnamefont {A.~M.}\ \bibnamefont {Archibald}}, \bibinfo {author}
  {\bibfnamefont {A.~H.}\ \bibnamefont {Ribeiro}}, \bibinfo {author} {\bibfnamefont {F.}~\bibnamefont {Pedregosa}}, \bibinfo {author} {\bibfnamefont {P.}~\bibnamefont {{van Mulbregt}}},\ and\ \bibinfo {author} {\bibnamefont {{SciPy 1.0 Contributors}}},\ }\bibfield  {title} {\enquote {\bibinfo {title} {{{SciPy} 1.0: Fundamental Algorithms for Scientific Computing in Python}},}\ }\href {https://doi.org/10.1038/s41592-019-0686-2} {\bibfield  {journal} {\bibinfo  {journal} {Nat. Methods}\ }\textbf {\bibinfo {volume} {17}},\ \bibinfo {pages} {261--272} (\bibinfo {year} {2020})}\BibitemShut {NoStop}%
\end{thebibliography}%

\end{document}